\documentclass[trackchanges,twocolumn]{aastex7}

\usepackage{url}  
\usepackage{hyperref}
\usepackage{graphicx}
\usepackage{amsmath}
\usepackage[normalem]{ulem}
\usepackage{xcolor}
\usepackage{tabularx}
\usepackage{graphicx}    
\usepackage{subcaption}  
\usepackage{xcolor}

\usepackage{float}  
\usepackage{booktabs}

\begin{document}

\title{
The Galaxy Luminosity Functions in ASTRID: Predictions for LSST
}

\author[orcid=0000-0002-9140-3950]{Fatemeh Hafezianzadeh}
\affiliation{McWilliams Center for Cosmology and Astrophysics, Department of Physics, Carnegie Mellon University, Pittsburgh, PA 15213, USA }
\email[show]{fhafezia@andrew.cmu.edu}  

\author[0000-0002-5596-198X]{Tianqing Zhang}
\affiliation{Department of Physics and Astronomy and PITT PACC, University of Pittsburgh, Pittsburgh, PA 15260, USA}\email{fakeemail2@google.com}
\author[0000-0002-1408-6904]{Paul Rogozenski}
\affiliation{McWilliams Center for Cosmology and Astrophysics, Department of Physics, Carnegie Mellon University, Pittsburgh, PA 15213, USA }\email{fakeemail2@google.com}
\author[0009-0006-7511-0329]{Patrick Lachance}
\affiliation{McWilliams Center for Cosmology and Astrophysics, Department of Physics, Carnegie Mellon University, Pittsburgh, PA 15213, USA }\email{fakeemail2@google.com}

\author[0000-0002-8828-8461]{Yihao Zhou}
\affiliation{McWilliams Center for Cosmology and Astrophysics, Department of Physics, Carnegie Mellon University, Pittsburgh, PA 15213, USA }\email{fakeemail2@google.com}

\author[orcid=0000-0002-6462-5734]{Tiziana Di Matteo} 
\affiliation{McWilliams Center for Cosmology and Astrophysics, Department of Physics, Carnegie Mellon University, Pittsburgh, PA 15213, USA }
\email{fakeemail2@google.com}

\author[orcid=0000-0003-0697-2583]{Rupert A. C. Croft} 
\affiliation{McWilliams Center for Cosmology and Astrophysics, Department of Physics, Carnegie Mellon University, Pittsburgh, PA 15213, USA }
\email{fakeemail2@google.com}

\author[orcid=0000-0001-5803-5490]{Simeon Bird} 
\affiliation{Department of Physics and Astronomy, University of California Riverside, 900 University Ave, Riverside, CA 92521, USA }
\email{fakeemail2@google.com}
\author[orcid=0000-0003-2271-1527]{Rachel Mandelbaum} 
\affiliation{McWilliams Center for Cosmology and Astrophysics, Department of Physics, Carnegie Mellon University, Pittsburgh, PA 15213, USA }
\email{fakeemail2@google.com}

\begin{abstract}
We present validated and forward-modelled galaxy luminosity functions and photometric predictions for the Vera C.\ Rubin Observatory Legacy Survey of Space and Time using the \textsc{ASTRID} cosmological hydrodynamical simulation. 
Galaxy magnitudes are computed by combining stellar population synthesis modeling with a physically motivated dust attenuation prescription in which the optical depth scales with metal surface density. 
The dust model is calibrated at $z=0$ using SDSS luminosity functions and tested at intermediate redshifts ($z=0.5$, $1.0$, and $1.5$) in rest-frame $B$, $V$, $R$, and $I$ bands. 
We find that the attenuated luminosity functions reproduce observed galaxy statistics across multiple wavelengths and redshifts.
Using this calibrated framework, we construct LSST-ready mock photometric catalogs over $0 \leq z \leq2$ in steps of $\Delta z=0.1$, containing $\sim$378 million galaxies.
We provide predicted apparent-magnitude luminosity functions in the LSST $ugrizy$ bands, derive best-fit Schechter parameters as a compact analytic representation, and compute differential and cumulative galaxy number counts as a function of survey depth from Year~1 to Year~10. 
\end{abstract}

\section{Introduction}

Galaxy luminosity functions (LFs) and broadband colors provide fundamental constraints on the physics of galaxy formation and evolution, including the buildup of stellar mass, the regulation of star formation, and the emergence of quenched populations across cosmic time \citep[e.g.,][]{Kauffmann2003,Baldry2006}. 
Because galaxies are observed through complex selection functions and are strongly affected by dust attenuation, connecting theoretical predictions to survey observables requires detailed forward modeling.

In particular, dust can significantly reshape both luminosity functions and color distributions, especially in optical and near-UV bands \citep[e.g.,][]{Calzetti2000,Pei1992}, making physically motivated attenuation models essential for reliable comparisons between simulations and data.

In addition to luminosities and colors, galaxy morphology provides a complementary probe of galaxy formation and evolution. In particular, the color--morphology relationship (CMR) connects observable galaxy properties to their underlying physical state, with early-type galaxies typically appearing red and spheroidal, and late-type galaxies appearing blue and disk-like \citep[see e.g.,][for reviews]{Conselice_2014, Somerville_2015}. 
This relation is especially important in the context of large-scale structure and weak lensing surveys, where recent studies aim to use this relationship to minimize systematic contamination from galaxy intrinsic alignments in cosmological inference \citep{10.1093/mnras/stz2197, 2024arXiv241022272M}.

The Vera C. Rubin Observatory Legacy Survey of Space and Time (LSST) will deliver a 10-year multi-band photometric survey for billions of galaxies across the entire southern sky, enabling unprecedented measurements of apparent-magnitude luminosity functions, number counts, and color--redshift relations out to $z \sim 2$ and beyond. 
However, because LSST observations are not yet available, robust theoretical predictions are necessary to anticipate survey yields, understand redshift-dependent selection effects, and provide validated mock catalogs for interpreting early data releases. 
Such predictions require simulations that simultaneously capture galaxy formation physics, large-scale structure, and the statistical abundance of both faint and rare massive systems.

Several mock galaxy catalogs and synthetic sky frameworks have been developed to support Rubin LSST science analyses using a variety of empirical, semi-analytic, and hybrid forward modeling approaches. The CosmoDC2, Buzzard, and Cardinal mock catalogs \citep{Korytov2019,DeRose2019,to2024buzzard} populate dark matter only $N$-body simulations using empirically calibrated galaxy halo connection models based on techniques such as abundance matching, halo occupation statistics, and observationally constrained galaxy properties. The Euclid Flagship simulations \citep{potter2017pkdgrav3,blanchard2020euclid} use large dark matter simulations together with semi-analytic and halo based prescriptions to construct realistic synthetic galaxy surveys, while OpenUniverse2024 \citep{openuniverse2025} provides a shared synthetic sky framework for Rubin and Roman analyses using empirical galaxy population models, observational calibrations, and forward survey simulations. These mock catalogs have become essential tools for survey validation, cosmological forecasts, covariance estimation, and pipeline development.However, most currently available LSST scale mocks do not directly evolve baryonic physics within a large volume hydrodynamical simulation. Since uncertainties in galaxy formation, dust attenuation, and feedback processes can introduce important systematic effects in mock based analyses, hydrodynamical simulations provide a highly complementary approach by self consistently connecting observable galaxy properties to the underlying evolution of gas, stars, metals, and black holes.

In this work, we construct LSST-like mock photometric catalogs from the \textsc{ASTRID} cosmological hydrodynamical simulation \citep{Bird2022,Ni2022_astrid,Ni2024}. 
Compared to other large-volume hydrodynamical simulations such as IllustrisTNG \citep{Pillepich2018, Nelson2018}, EAGLE \citep{Schaye2015}, and SIMBA \citep{Dave2019}, ASTRID offers a particularly advantageous combination of large volume and high mass resolution. 
With a periodic box of $(250\,h^{-1}\mathrm{cMpc})^3$ and $2\times5500^3$ particles, ASTRID resolves galaxy populations over a wide stellar mass range while simultaneously sampling rare massive halos relevant for wide-field surveys like LSST. 
In addition, ASTRID includes an improved treatment of massive black hole dynamics based on dynamical friction rather than repositioning schemes, leading to more realistic black hole trajectories and feedback histories \citep{Bird2022,Ni2022_astrid}. 
The simulation has been shown to reproduce key low-redshift observables, including the stellar mass function, galaxy luminosity functions (when dust is included), and color bimodality, while also hosting a statistically significant population of massive clusters at $z=0$ \citep{zhou2026astrid}. 
This combination of validated galaxy statistics, physical feedback modeling, and survey-scale volume makes ASTRID particularly well suited for constructing predictive LSST mock catalogs over $0<z<2$.

To generate observable quantities, we compute intrinsic stellar emission for each galaxy using the \textsc{FSPS} stellar population synthesis model \citep{Conroy2009,Conroy2010}, applied to individual star particles treated as simple stellar populations. 
We then implement a physically motivated dust attenuation model in which the optical depth scales with the line-of-sight metal surface density and follows a power-law wavelength dependence. 
The attenuation model contains two free parameters controlling its normalization and wavelength slope, which we calibrate at low redshift using observed luminosity functions and then test for consistency at higher redshift.

After validating the predicted luminosity functions and color distributions against existing surveys, we present forward predictions for LSST in the $ugrizy$ bands across $0<z \leq 2$. 
These predictions include apparent-magnitude luminosity functions, best-fit Schechter parameters, differential and cumulative number counts as a function of survey depth, the evolution of color distributions with redshift, and galaxy morphology for redshift zero.
Together, these results provide a physically motivated and observationally calibrated framework for interpreting future LSST galaxy samples.

This paper is organized as follows. 
Section~\ref{sec:methods} describes the \textsc{ASTRID} simulation, the intrinsic luminosity modeling, and the dust attenuation prescription. 
Section~\ref{sec:validation} validates the model against low-redshift luminosity functions and color--color distributions and tests the redshift evolution of the LF. 
Section~\ref{sec:prediction} presents LSST forward predictions for apparent-magnitude LFs, Schechter parameters, number counts, and color evolution. 
We summarize the main results and discuss limitations and future improvements in Section~\ref{sec:discussion}.

\section{method}
\label{sec:methods}

\subsection{ASTRID Simulation}
We use the \textsc{ASTRID} \citep{Bird2022,Ni2022_astrid,Ni2024} cosmological hydrodynamical simulation as the theoretical baseline for our analysis. 
ASTRID was run with the Smoothed-Particle Hydrodynamics (SPH) code \textsc{MP-Gadget} \citep{Feng2018} and evolves a periodic volume of side length $250\,h^{-1}\mathrm{cMpc}$ from initial conditions set at $z=99$, using $2\times 5500^3$ tracer particles representing dark matter and baryons. 
The simulation adopts a Planck cosmology \citep{planck2015} with $h=0.6774$, $\Omega_{\mathrm{m}}=0.3089$, $\Omega_{\Lambda}=0.6911$, $\Omega_{\mathrm{b}}=0.0486$, $\sigma_8=0.816$, and $n_s=0.9667$. 
The corresponding mass resolution is $M_{\rm DM}=6.7\times10^6\,h^{-1}M_\odot$ for dark matter and $M_{\rm gas}=1.3\times10^6\,h^{-1}M_\odot$ for gas, with a (comoving) gravitational softening length of $\epsilon_g=1.5\,h^{-1}\mathrm{ckpc}$ for both components. 
ASTRID includes a full-physics subgrid model suite to capture unresolved galaxy-formation processes, including prescriptions for massive black holes, stellar and AGN feedback, and inhomogeneous hydrogen and helium reionization, enabling self-consistent predictions for galaxy and black-hole populations at low redshift \citep{zhou2026astrid}.

\subsection{Intrinsic Galaxy Luminosities}
\label{subsec:intrinsic_lf}
Galaxy luminosity functions and galaxy colors are key observables for constraining models of star formation and stellar evolution \citep{Kauffmann2003, Baldry2006}. 
To model the intrinsic (i.e., dust-free) galaxy luminosities in ASTRID, we compute the stellar emission associated with each galaxy by summing the contributions from its constituent star particles.

Each star particle is treated as a simple stellar population (SSP), characterized by its formation time, metallicity, and stellar mass as recorded in the simulation. 
We model the stellar emission using the \textsc{FSPS} stellar population synthesis code \citep{Conroy2009, Conroy2010}, adopting the PARSEC isochrones \citep{Bressan2012_parsec} and the MILES stellar library \citep{Sanchez-Blazquez2006_miles}. 
A Chabrier initial mass function (IMF) is assumed throughout \citep{Chabrier2003}. 
The intrinsic luminosity of a galaxy in a given photometric band is obtained by combining the luminosity contributions of all star particles gravitationally bound to that galaxy.

Rather than computing full SEDs for each star particle at runtime, we adopt a computationally efficient, grid-based implementation of \textsc{FSPS}. 
This approach is commonly used in large-volume simulation analyses to reduce the computational cost associated with assigning stellar magnitudes to tens of millions of star particles \citep[e.g.,][]{Pan2023}. 
Specifically, we precompute intrinsic, dust-free SSP magnitudes for each photometric band on a two-dimensional grid of stellar age and metallicity. 
For each grid point, magnitudes are calculated using \textsc{FSPS}, and individual star particle magnitudes are subsequently obtained by interpolating within this age--metallicity grid according to the particle’s age and metallicity and scaling by its stellar mass. 
Particle luminosities are then summed to obtain the total intrinsic luminosity of each galaxy, which is finally converted back to magnitudes.

This grid-based method is mathematically equivalent to deriving magnitudes by integrating SSP spectral energy distributions through the corresponding filter transmission curves, since both approaches rely on the same underlying stellar population synthesis physics. 
Comparisons between magnitudes derived using different SPS implementations and methodologies have shown that grid-based and SED-based magnitudes typically agree to within $\lesssim 0.1$ dex, with larger discrepancies arising primarily in metal-rich systems or populations dominated by young and intermediate-age stars, where SPS model uncertainties are largest \citep{Guidi2015}.

The intrinsic luminosity model described here therefore provides an accurate and robust baseline representation of the stellar emission in ASTRID, independent of dust. 
This baseline serves as the reference for applying dust attenuation models and for constructing mock photometric catalogs designed to emulate LSST observations.

\subsection{Dust-Attenuated Galaxy Luminosities}
\label{subsec:attenuated_lf}

To model dust attenuation, we attenuate the stellar emission of each star particle based on the distribution of metals in the interstellar medium (ISM) along its line of sight. 
Following previous work \citep{Wilkins2017, lachance2024}, we assume that the dust optical depth scales with the metal surface density and follows a wavelength-dependent power-law attenuation curve.

For each galaxy, we construct a three-dimensional metal density field using the gas particles associated with the system. 
The metal surface density, $\Sigma(x,y,z)$, is computed by integrating the metal density along the line of sight in front of each star particle. 
The wavelength-dependent optical depth is then given by
\begin{equation}
\tau_{\mathrm{ISM}}(\lambda) = \kappa_{\mathrm{ISM}} \, \Sigma(x,y,z)
\left(\frac{\lambda}{0.55\,\mu\mathrm{m}}\right)^{\gamma},
\end{equation}
where $\kappa_{\mathrm{ISM}}$ sets the overall normalization of the attenuation and is independent of wavelength, while $\gamma$ controls the wavelength dependence of the attenuation curve. 
This functional form yields an attenuation law that is steeper than the Calzetti starburst law \citep{Calzetti2000} but flatter than the Small Magellanic Cloud curve \citep{Pei1992} for appropriate choices of $\gamma$.

The dust attenuation model therefore contains two free parameters, $\kappa_{\mathrm{ISM}}$ and $\gamma$, which must be calibrated simultaneously to provide a consistent description of galaxy luminosities across different wavelength bands. 
Since $\kappa_{\mathrm{ISM}}$ controls the overall normalization of the attenuation and is independent of wavelength, we first determine its optimal value using the SDSS $g$ band, which is centered near $0.55\,\mu$m and is therefore minimally affected by the wavelength-dependent term. 
We explore values of $\kappa_{\mathrm{ISM}}$ in the range $2 \leq \kappa_{\mathrm{ISM}} \leq 4$ and identify the best-fitting value by minimizing the $\chi^2$ statistic between the simulated and observed $g$-band luminosity functions, adopting $\kappa_{\mathrm{ISM}} = 2.9$ as the fiducial normalization.

With $\kappa_{\mathrm{ISM}}$ fixed, we then calibrate the wavelength dependence of the attenuation by varying $\gamma$ over the range $-0.5$ to $-1.5$. 
For each value of $\gamma$, we compute dust-attenuated luminosity functions in multiple photometric bands and evaluate the combined $\chi^2$ across bands to identify a single value of $\gamma$ that provides a consistent fit over a wide wavelength range. 
We find that $\gamma = -0.7$ yields the best overall agreement across different bandpasses.

The calibration of both parameters is first performed at $z=0$, where high-quality observational constraints are available. 
We then test the resulting dust model at higher redshifts ($z=0.5$, $1.0$, and $1.5$) using rest-frame $B$, $V$, $R$, and $I$ bands to assess its robustness and redshift independence. 
The results of these tests are presented in Section~\ref{sec:Redshift Evolution}.

After verifying that the dust-attenuated luminosity functions obtained with this model are consistent with observational measurements over a range of wavelengths and redshifts, we adopt this fiducial dust attenuation model to compute dust-attenuated magnitudes for the LSST mock galaxy catalogs used throughout this work.

\subsection{Galaxy Morphology}
\label{subsec:morphology}

To characterize galaxy morphology in ASTRID, we determine galaxy shapes using the stellar mass distribution of bound star particles within each subhalo.
We calculate the simple inertia tensor, which has been shown to produce non-zero alignment signals in hydrodynamical simulations and ellipticity distributions comparable to observations \citep{10.1093/mnras/stac1424}. For $n$ stellar particles with mass $m_n$ at positions $r_n$ relative to the galaxy center, the inertia tensor is defined as

\begin{equation}
I_{ij} = \frac{\sum_n m_n r_{ni} r_{nj}}{\sum_n m_n}.
\end{equation}

We evaluate the eigenvalues of the inertia tensor and take the principal axes of the resulting ellipsoid to be proportional to the square root of the eigenvalues. The principal axes are defined such that $a \geq b \geq c$, and we use axis ratios (e.g., $c/a$) as a measure of galaxy morphology. 
We follow the formalism of \citet{10.1093/mnras/stac1424,2013MNRAS.431..477J} to define projected ellipticities, which we denote as $|e|$.

To connect morphology with observable galaxy properties, we define red and blue galaxy populations at $z=0$ for galaxies with a stellar mass threshold of $M_* \geq 10^{9.5} M_\odot$ using a color--magnitude split in $M_g-M_r$ and $M_r$ space, following a similar approach to recent studies \citep{10.1093/mnras/staa3802,2025AA...699A.252G}. 
We adopt the dividing line

\begin{equation}
M_g-M_r = 0.30 - 0.015 M_r,
\end{equation}

such that galaxies above this line are classified as red and those below as blue.

\section{Results}
In this section, we validate the predicted LSST galaxy magnitudes and colors using currently available observational data (Section~\ref{sec:validation}). 
Since observations with the LSST Camera are not yet available at the time of the publication, other than a small commissioning dataset \citep{rtn-095}, a direct comparison between our mock LSST catalogs and survey data is not possible. 
We therefore first compare our model predictions with existing optical and near infrared measurement of luminosity functions and color distributions to establish the reliability of the galaxy population and dust treatment. 
In Section~\ref{sec:prediction}, we then present forward predictions for LSST apparent magnitudes and number counts based on the validated model.

\subsection{Validation}
\label{sec:validation}
To validate the predicted galaxy magnitudes, we evaluate the simulation in several ways. 
First, in Section~\ref{sec:Low redshift validation}, we compare the model predictions for galaxies at $z<0.1$ across multiple bands in order to calibrate and determine the best-fit dust attenuation parameters. 
Next, in Section~\ref{sec:Redshift Evolution}, we demonstrate that the tuned attenuation model provides good agreement with observations over a range of redshifts. 
Finally, in Section~\ref{sec:Color-Color}, we assess the color--color distributions of galaxies at $z<0.1$ as an additional validation of the model.
\begin{figure*}[ht]
\centering
\includegraphics[width=\textwidth , trim=0 9cm 0 0, clip]{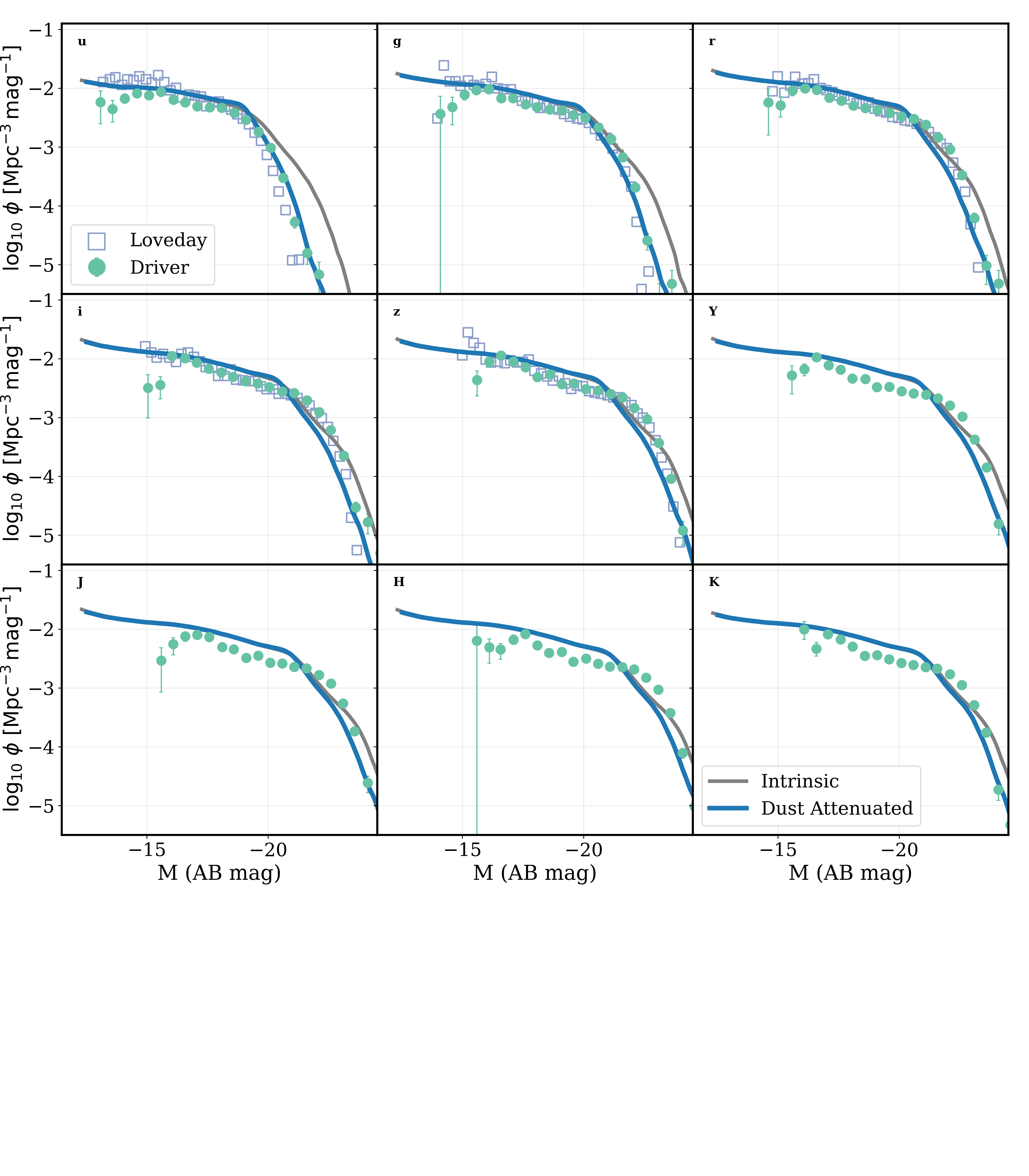}
\caption{Galaxy luminosity functions at $z=0$ in the SDSS $u,g,r,i,z$ and near-infrared $Y,J,H,K$ bands.
Gray curves show intrinsic ASTRID predictions, while blue curves show dust-attenuated luminosity functions.
Observational measurements from Loveday et al.\ (open squares) and Driver et al.\ (filled circles) are overplotted.}
    
\label{fig:LF_z=0}
\end{figure*}
\label{sec:validation}

\begin{figure}
    \centering
    \includegraphics[width=0.45\textwidth]{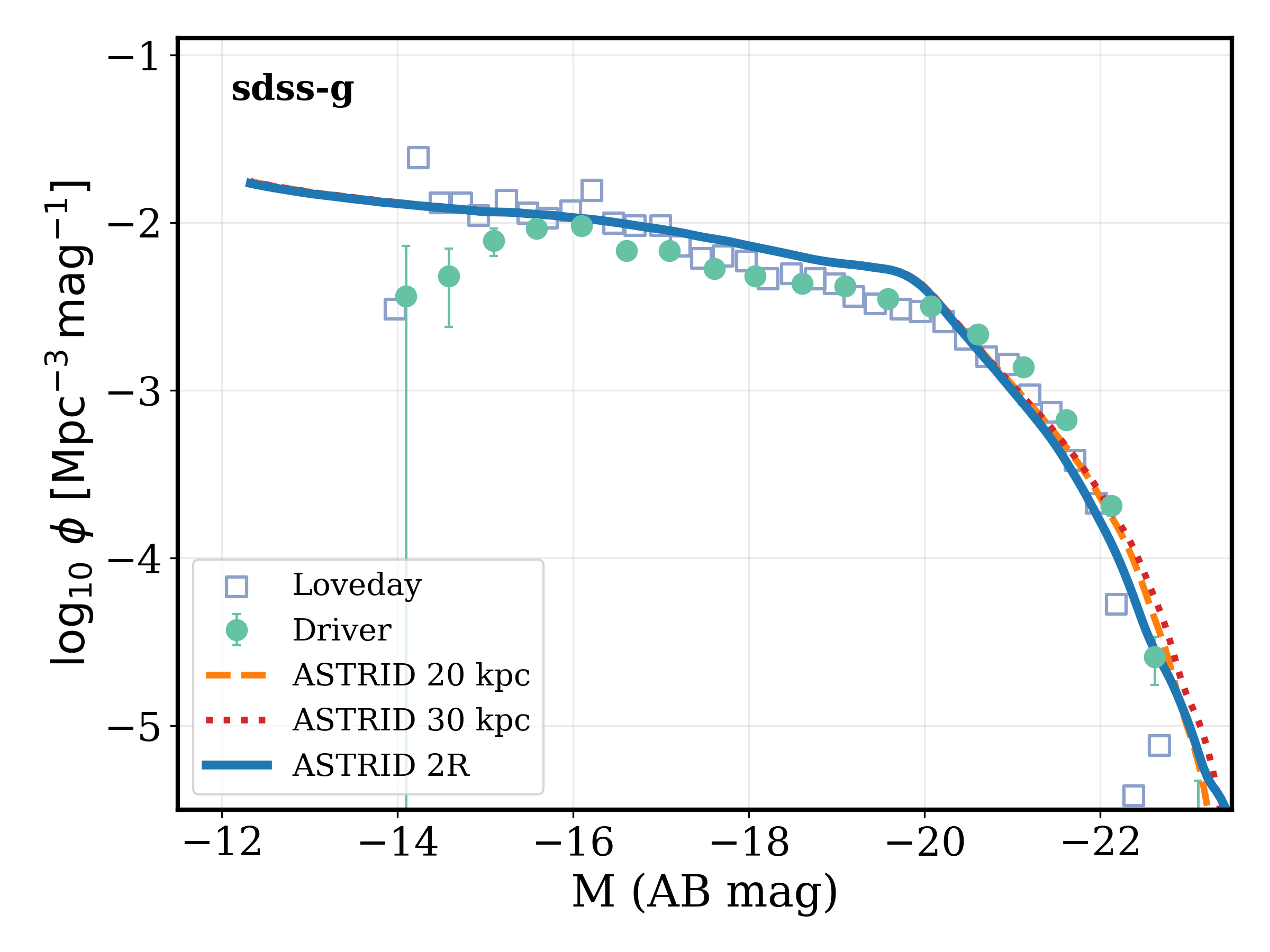}
    \caption{
    SDSS-$g$ band luminosity function at $z=0$ comparing observational measurements
    from \citet{loveday2012galaxy} and \citet{driver2012galaxy} with ASTRID predictions computed using
    different photometric apertures: 20~kpc, 30~kpc, and $2R_{\rm half}$.
    }
    \label{fig:LF_z0_aperture}
\end{figure}
\begin{figure*}[ht!]
\centering
\includegraphics[width=0.9\textwidth , trim=2cm 2cm 0 0, clip]{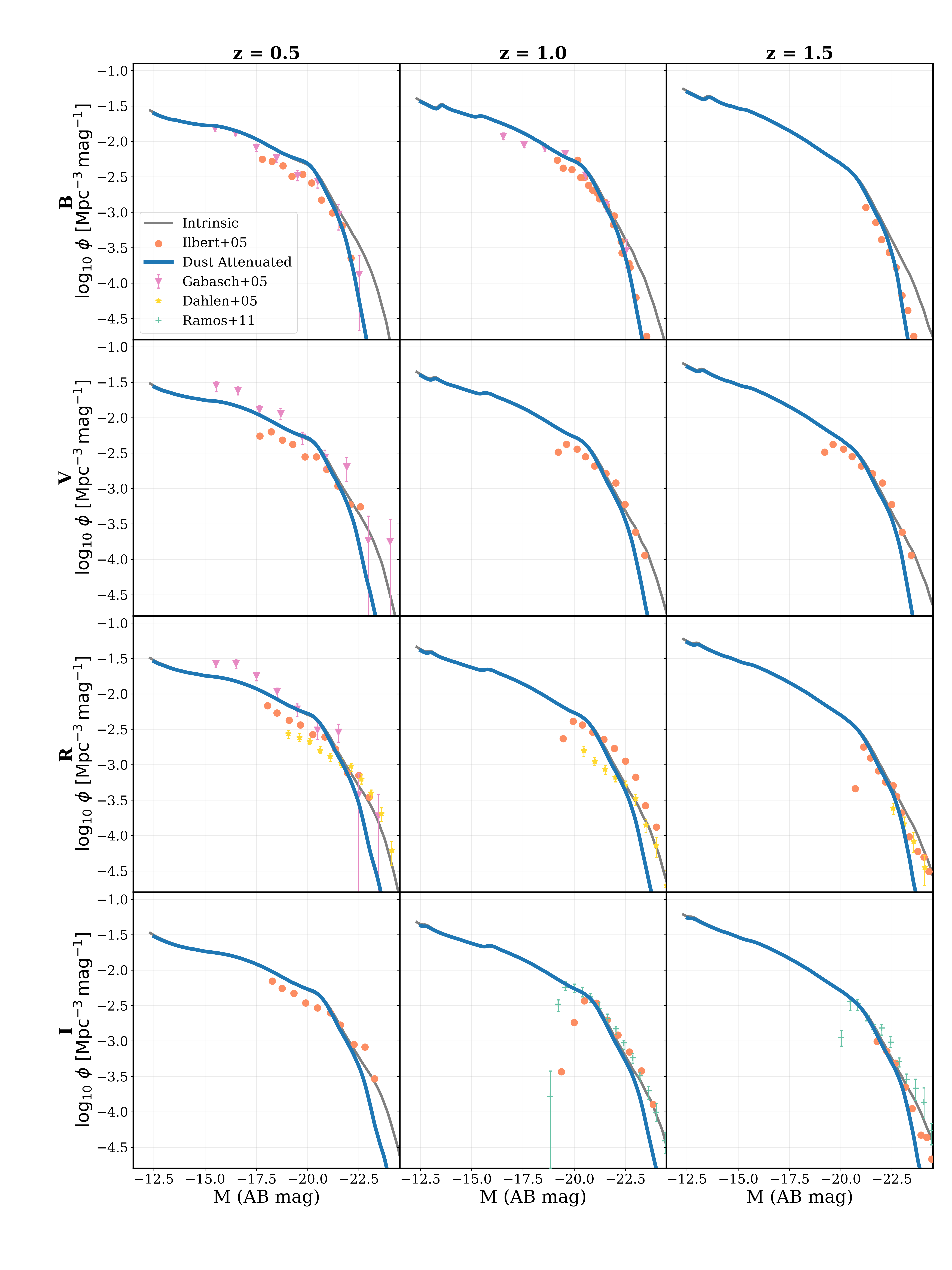}
\caption{Rest-frame $B$, $V$, $R$, and $I$ luminosity functions at $z=0.5$, $1.0$, and $1.5$. 
Gray curves show intrinsic simulation predictions, while blue curves include dust attenuation. 
Observational measurements from \citet{Ilbert2005}, \citet{Gabasch2004}, \citet{Dahlen2005}, and \citet{Ramos2011} are overplotted for comparison.}
    
\label{fig:LF_redshift_evolution}
\end{figure*}
\subsubsection{Low redshift validation}
\label{sec:Low redshift validation}

Figure~\ref{fig:LF_z=0} compares the intrinsic and dust-attenuated galaxy luminosity functions predicted by ASTRID at $z=0$ with observational measurements across optical and near-infrared bands. 
All ASTRID luminosity functions shown in this figure, as well as in the following figures, are computed using galaxy magnitudes measured within an aperture of twice the stellar half-mass radius ($2R_{\rm half}$).

Dust attenuation systematically reduces the number density of bright galaxies, with the strongest suppression occurring in the bluer SDSS bands and a progressively weaker impact toward longer wavelengths. 
At the bright end ($M \le -21$), the mean offset between the dust-attenuated and intrinsic luminosity functions,
$\Delta\log_{10}\phi \equiv \log_{10}\phi_{\rm att}-\log_{10}\phi_{\rm int}$, is $\langle\Delta\log_{10}\phi\rangle \simeq -0.98$ in $u$ and $-0.68$ in $g$, decreasing to $-0.50$, $-0.43$, and $-0.36$ in $r$, $i$, and $z$, respectively. 
In the near-infrared bands ($YJHK$), the bright-end suppression is significantly smaller, with $\langle\Delta\log_{10}\phi\rangle \simeq -0.31$, $-0.28$, $-0.22$, and $-0.24$ for $Y$, $J$, $H$, and $K$, consistent with the reduced sensitivity of longer wavelengths to dust extinction.

Across all bands, the dust-attenuated ASTRID luminosity functions closely reproduce the observational measurements over the magnitude range directly probed by the data, supporting the realism of both the simulated galaxy population and the adopted dust prescription at $z=0$. 

Importantly, ASTRID also predicts the faint-end behavior beyond current observational limits. 
The observational compilations reach faint-end limits of $M^{\rm obs}_{\rm faint} \simeq -13.1$ in $u$, $-14.0$ in $g$, and $\sim -14.6$ to $-16.1$ in $rizYJHK$, whereas the simulation extends uniformly to $M^{\rm sim}_{\rm faint} \simeq -12.52$ in all bands. 
This corresponds to an additional $\Delta M_{\rm faint} \simeq 0.6$--$3.6$ magnitudes of faint-end coverage, depending on band. 
At this simulated faint limit, the predicted number density remains substantial, with $\log_{10}\phi(M^{\rm sim}_{\rm faint}) \simeq -1.7$ to $-1.9~{\rm Mpc^{-3}\,mag^{-1}}$, indicating a significant population of low-luminosity galaxies that are currently below observational detection thresholds.

To quantify potential systematic effects associated with the photometric aperture definition, we further examine the SDSS-$g$ band luminosity function at $z=0$ using dust-attenuated galaxy magnitudes measured within three different apertures: fixed physical radii of 20~kpc and 30~kpc, and an adaptive aperture defined as $2R_{\rm half}$ \citep{Donnari2019}. 
As shown in Figure~\ref{fig:LF_z0_aperture}, the resulting dust-attenuated luminosity functions are very similar across the full magnitude range. 
At the bright end ($M \le -21$), the magnitude offsets relative to the $2R_{\rm half}$ aperture remain modest: $\Delta M \simeq -0.17$ to $-0.13$~mag for the 20~kpc aperture and $\Delta M \simeq -0.23$ to $-0.12$~mag for the 30~kpc aperture at number densities $\phi = 10^{-4}$--$10^{-5}\,{\rm Mpc^{-3}\,mag^{-1}}$. 
These differences are small compared to the observational uncertainties and do not significantly affect the shape or normalization of the dust-attenuated luminosity function. 
We therefore adopt the adaptive aperture of $2R_{\rm half}$ for the remainder of this work, which is also conceptually consistent with LSST model-based photometry (e.g., CMODEL), where adaptive measurements better capture extended galaxy light profiles.

\begin{figure*}[ht]
\centering
\includegraphics[width=\textwidth]{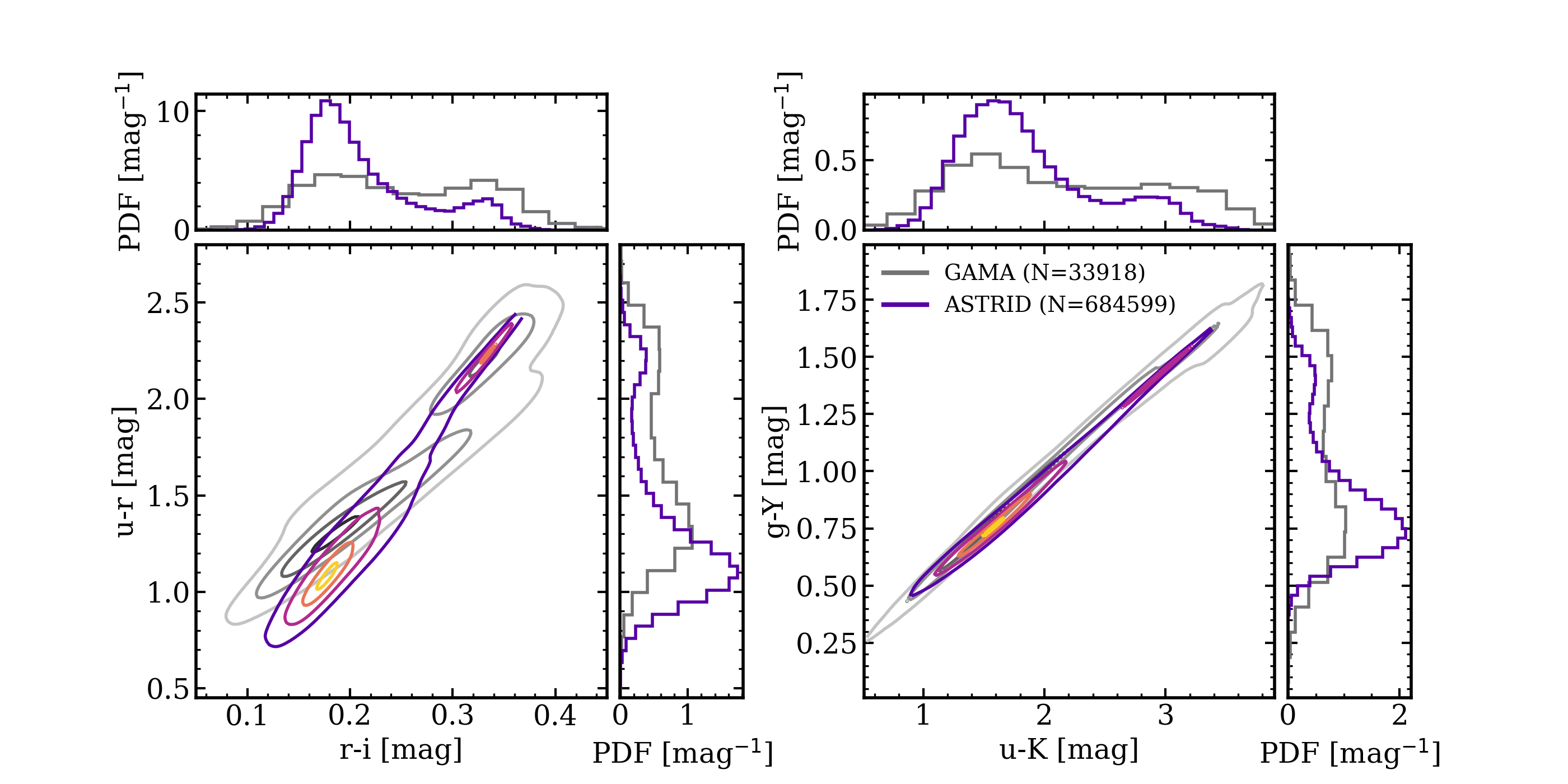}
\caption{Four different color--color relations comparing the simulation (purple) and observational data from the GAMA survey (gray).
All galaxies are selected in the redshift range $0.002 \le z < 0.1$ and with stellar masses $M_\star \ge 10^{8.5}\,M_\odot$.
The number of galaxies used in each panel is indicated in the top-right corner.
The two-dimensional distributions are estimated using a kernel density estimate (KDE);
contours correspond to enclosed fractions of 5, 25, 60, and 90\% of the total kernel density.
One-dimensional marginal probability density functions for each color are shown along the top and right axes.}
    
\label{fig:valid_color_color}
\end{figure*}

\subsubsection{Redshift Evolution}
\label{sec:Redshift Evolution}
Figure~\ref{fig:LF_redshift_evolution} provides an additional validation of the ASTRID LF predictions beyond the local Universe. We present the intrinsic and dust-attenuated rest-frame $B$, $V$, $R$, and $I$ luminosity functions at $z=0.5$, $1.0$, and $z=1.5$, compared with observational measurements from \citet{Ilbert2005}, \citet{Gabasch2004}, \citet{Dahlen2005}, and \citet{Ramos2011}. 

As expected, the intrinsic LFs systematically overpredict the observed number densities, particularly toward the bright end, reflecting the absence of dust extinction. After applying our attenuation model, the predicted LFs shift to fainter magnitudes and achieve substantially improved agreement with the data. The effect of dust is most pronounced in the $B$ band, where attenuation significantly suppresses the intrinsic luminosities. In this band, the attenuated LF shows excellent agreement with the observational measurements across the full magnitude range probed, reproducing both the normalization and the overall shape within the reported uncertainties. 

In the redder $V$, $R$, and $I$ bands, the impact of dust is comparatively weaker, consistent with the wavelength dependence of extinction. Although the difference between intrinsic and attenuated LFs is smaller in these bands, the attenuated predictions systematically lie within the observational error bars. 

Overall, the agreement between the attenuated model and the observational data across multiple bands and redshifts demonstrates that our dust prescription provides a physically consistent description of galaxy luminosities at intermediate redshifts. Together with the $z=0$ validation, these results indicate that ASTRID successfully reproduces the observed evolution of galaxy luminosity functions.

\subsubsection{Color-Color Validation}
\label{sec:Color-Color}

For Figure~\ref{fig:valid_color_color}, we use the GAMA DR4 (v24) \texttt{StellarMasses} table to obtain stellar masses and rest-frame colors for the observational sample. 
We select galaxies in the redshift range $0.002 \leq z < 0.1$ and require $M_\star > 10^{8.5}\,M_\odot$, based on the mass distribution and completeness limits of the data. 
For the simulation, we use the $z=0$ snapshot of the ASTRID simulation and apply the same stellar-mass cut. 
To ensure that only well-resolved systems are included, we additionally require the stellar half-mass radius to satisfy $R_{1/2,\star} > 2~\mathrm{kpc}/h$. 
To mimic the main GAMA selection, we further impose an apparent magnitude cut of $m_r < 19.8$ on the simulated catalog. 

The specific star formation rate is defined as $\mathrm{sSFR} = \mathrm{SFR}/M_\star$, where both the star formation rate and stellar mass are measured within the stellar half-mass radius of each galaxy. 
Finally, to avoid including extremely star-forming systems, we apply a cut of $\log_{10}(\mathrm{sSFR}) < -9.8$ to the simulated galaxies.
In each panel, the gray contours show the GAMA distribution and the colored contours show the simulation. 
Contours correspond to enclosed fractions of 5, 25, 60, and 90\% of the total two–dimensional kernel density estimate (KDE). 
One–dimensional marginal probability density functions (PDFs) are shown along the top and right axes for each color.

Overall, ASTRID successfully reproduces the bimodality of the galaxy population, corresponding to the observed blue cloud and red sequence. In several color combinations—such as $u-K$, $g-Y$, and $r-i$—the agreement with observations is very good. However, in some cases (e.g., $u-r$) a small systematic shift toward slightly bluer colors is present. 

For example, in $u-r$ we find two clear peaks: the blue peak occurs at $u-r \simeq 1.1$ in the simulation compared to $u-r \simeq 1.3$ in GAMA, while the red peak appears at $u-r \simeq 2.2$ in the simulation compared to $u-r \simeq 2.3$ in the observations. These peak locations are broadly consistent with observations, although the blue sequence is slightly shifted toward bluer values in ASTRID. The blue peak is dominated by star-forming galaxies with young stellar populations, while the red peak corresponds to quiescent systems with suppressed star formation.

This shift can have several possible explanations. First, stellar masses in the observational data are subject to uncertainties, whereas ASTRID contains a significant population of low-mass galaxies with relatively high sSFR, which can bias the simulated color distribution toward bluer values. Consistent with this interpretation, applying a higher stellar-mass threshold in ASTRID shifts the blue peak toward redder colors. 

Second, the adopted dust attenuation model is relatively simple; more sophisticated prescriptions—such as those including stellar age dependence—could potentially yield colors in closer agreement with observations. 

Finally, ASTRID predicts a relatively high star formation rate density at $z=0$ compared to some observational estimates (see Fig.~14 of \citealt{zhou2026astrid}). This may indicate an overproduction of highly star-forming galaxies at low redshift, naturally leading to a larger fraction of blue galaxies in the simulation.

\subsection{LSST Prediction}
\label{sec:prediction}
\begin{figure*}[ht]
    \centering
    \includegraphics[
        width=1.1\textwidth,
        trim=1.1cm 0cm 0cm 0cm,
        clip
    ]{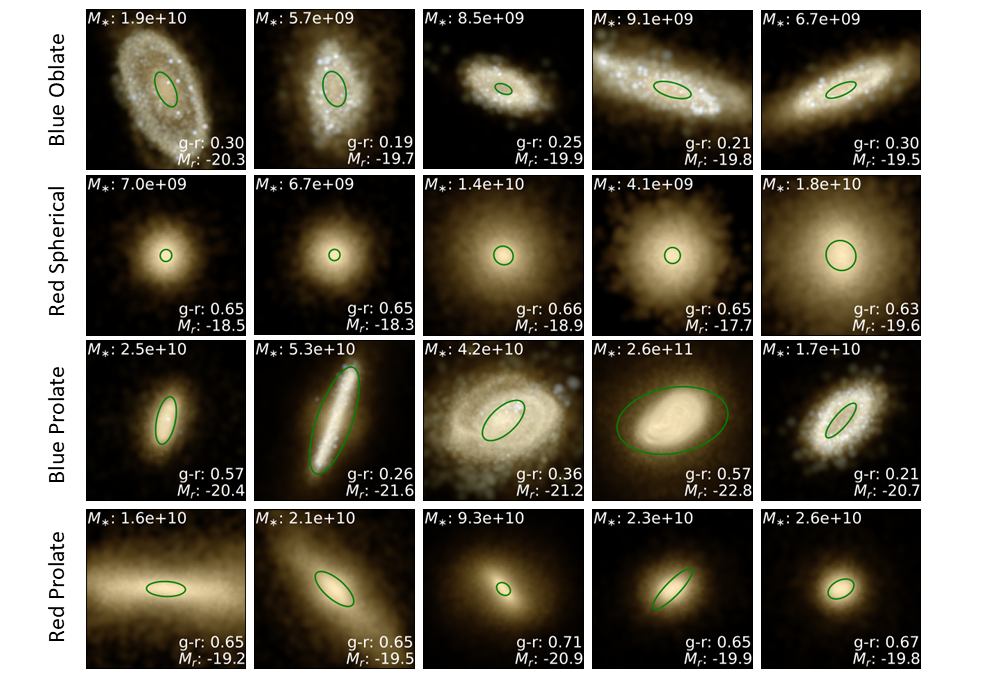}
    
    \caption{Example LSST-like mock galaxy images from the ASTRID simulation. From top to bottom: blue oblate (disk-like), red spherical, blue prolate, and red prolate galaxies. Green ellipses show the projected galaxy shapes. Each panel also lists the stellar mass ($M_\ast$), color ($g-r$), and $r$-band absolute magnitude ($M_r$).}
    
    \label{fig:image_grid}
\end{figure*}

Our LSST mock catalog spans the redshift range $0 \leq z \leq 2$ in steps of $\Delta z = 0.1$, enabling predictions for luminosity functions, galaxy number densities, projected shapes, color evolution, and LSST-like galaxy images. Figure~\ref{fig:image_grid} shows example mock observations of galaxies with different colors and morphologies, including blue oblate (disk-like), red spherical, blue prolate, and red prolate systems. The images were generated following \citet{lachance2024}, assuming LSST resolution and a seeing PSF of 0.4 arcsec.

\begin{table*}[t]
\centering
\caption{Median coadded depth per year for LSST in the six photometric bands, in AB magnitudes, showing both the 5$\sigma$ point-source (PS) depth from the simulated survey cadence\footnote{\url{https://github.com/yoachim/25_scratch/blob/main/depth_at_time/vector_depth.ipynb}} and the 5$\sigma$ extended-source (EX) limiting magnitude derived from our photometric error model. Extended-source limits are $\sim$0.3--0.5\,mag shallower than point-source limits due to the additional sky background subtended by the galaxy profile.}
\label{tab:lsst_coadd_depth}
\begin{tabular}{c|c|c|c|c|c|c|c|c|c|c|c|c}
\hline
Year
  & $u_\mathrm{PS}$ & $u_\mathrm{EX}$
  & $g_\mathrm{PS}$ & $g_\mathrm{EX}$
  & $r_\mathrm{PS}$ & $r_\mathrm{EX}$
  & $i_\mathrm{PS}$ & $i_\mathrm{EX}$
  & $z_\mathrm{PS}$ & $z_\mathrm{EX}$
  & $y_\mathrm{PS}$ & $y_\mathrm{EX}$ \\
\hline
 1 & 24.44 & 24.10 & 25.60 & 25.20 & 25.59 & 25.28 & 25.17 & 24.70 & 24.47 & 23.98 & 23.56 & 23.04 \\
 2 & 24.65 & 24.48 & 25.97 & 25.59 & 26.03 & 25.66 & 25.53 & 25.08 & 24.82 & 24.36 & 23.91 & 23.42 \\
 3 & 24.84 & 24.70 & 26.15 & 25.81 & 26.26 & 25.88 & 25.76 & 25.30 & 25.07 & 24.59 & 24.16 & 23.64 \\
 4 & 24.98 & 24.86 & 26.25 & 25.97 & 26.39 & 26.04 & 25.88 & 25.46 & 25.22 & 24.74 & 24.29 & 23.80 \\
 5 & 25.12 & 24.98 & 26.36 & 26.09 & 26.50 & 26.16 & 25.98 & 25.58 & 25.33 & 24.87 & 24.40 & 23.92 \\
 6 & 25.23 & 25.08 & 26.45 & 26.19 & 26.59 & 26.26 & 26.07 & 25.68 & 25.43 & 24.97 & 24.49 & 24.02 \\
 7 & 25.33 & 25.17 & 26.54 & 26.27 & 26.68 & 26.34 & 26.16 & 25.76 & 25.52 & 25.05 & 24.57 & 24.11 \\
 8 & 25.39 & 25.24 & 26.61 & 26.35 & 26.74 & 26.42 & 26.22 & 25.84 & 25.58 & 25.13 & 24.63 & 24.18 \\
 9 & 25.46 & 25.31 & 26.69 & 26.41 & 26.81 & 26.48 & 26.29 & 25.90 & 25.65 & 25.19 & 24.70 & 24.24 \\
10 & 25.53 & 25.36 & 26.76 & 26.47 & 26.88 & 26.54 & 26.37 & 25.96 & 25.72 & 25.25 & 24.76 & 24.30 \\
\hline
\end{tabular}
\end{table*}

\subsubsection{Luminosity Function}
From these catalogs, we construct apparent-magnitude LFs in all six LSST photometric bands and derive the corresponding Schechter parameters, enabling direct analytical access to the LF at each redshift. 
While the full redshift grid is used in the analysis, we display a representative subset of redshifts to highlight the main trends in redshift and color evolution.

\begin{figure*}[ht]
    \centering
    \includegraphics[width=0.32\textwidth]{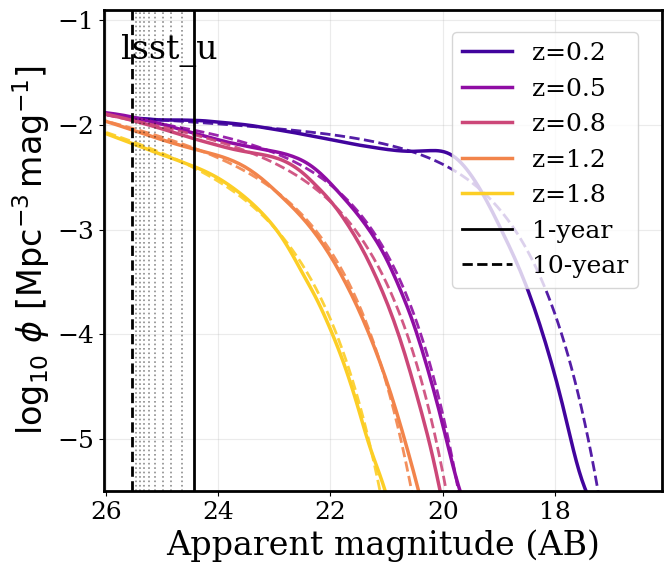}
    \includegraphics[width=0.32\textwidth]{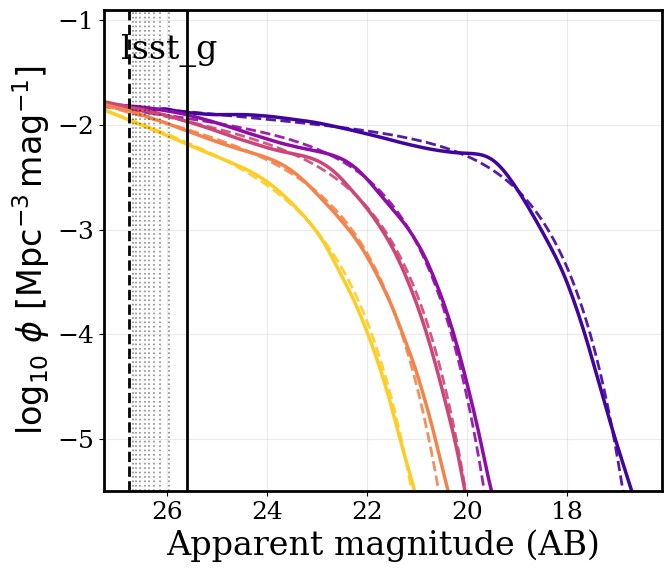}
    \includegraphics[width=0.32\textwidth]{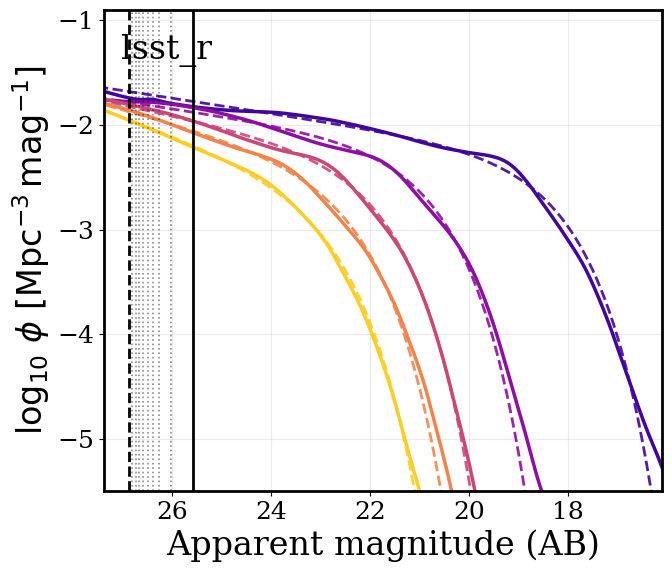}

    \includegraphics[width=0.32\textwidth]{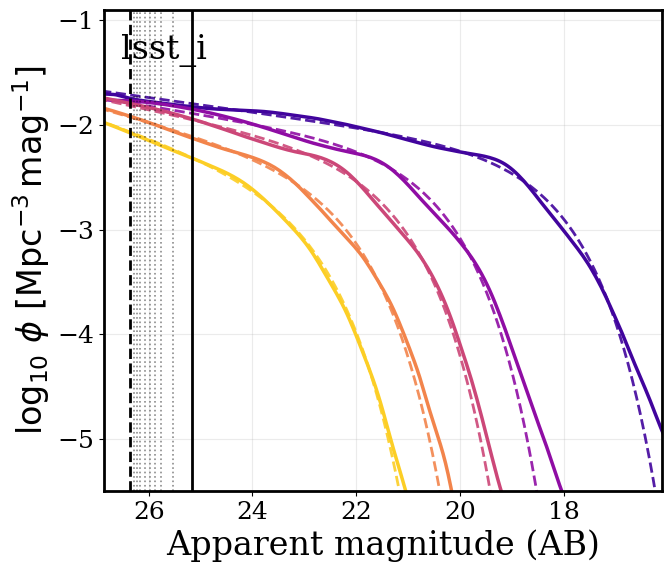}
    \includegraphics[width=0.32\textwidth]{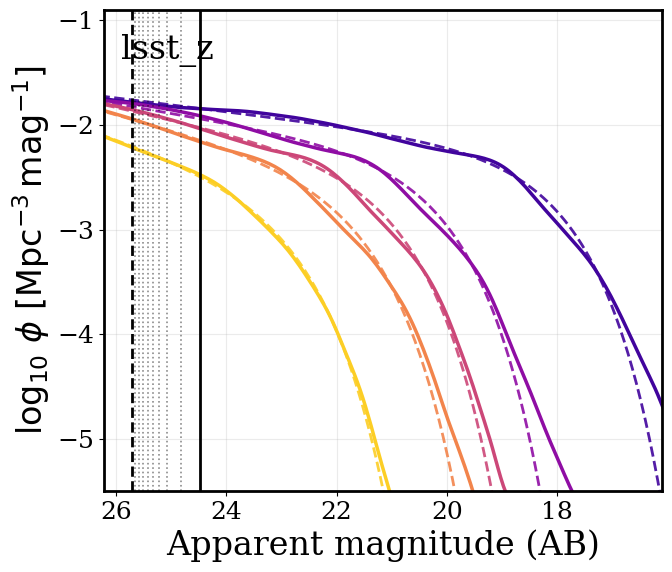}
    \includegraphics[width=0.32\textwidth]{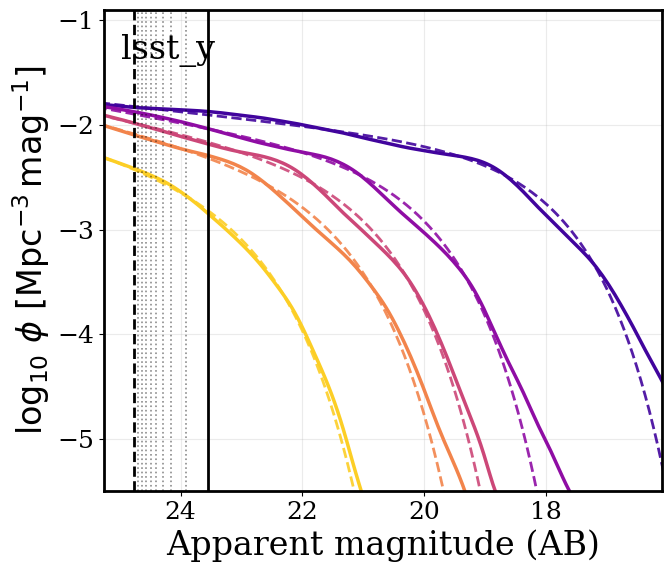}
    \caption{Apparent-magnitude number densities in the LSST $ugrizy$ bands. Each panel shows the number density of galaxies as a function of observed magnitude at multiple redshifts. Vertical lines indicate the LSST $5\sigma$ limiting depths: the solid line corresponds to the first-year coadded depth, dotted lines denote intermediate coadded depths from two to nine years, and the dashed line marks the 10-year coadded depth.}
    \label{fig:lsst_nd_3x2}
\end{figure*}

Figure~\ref{fig:lsst_nd_3x2} presents the predicted apparent-magnitude galaxy number densities in the LSST $u$, $g$, $r$, $i$, $z$, and $y$ bands at $z=0.2$, $0.5$, $0.8$, $1.2$, and $1.8$. 
In each panel, solid colored curves show the luminosity functions directly measured from our mock catalogs, while dashed curves of the same color indicate the corresponding Schechter fits. 
Darker to lighter colors represent increasing redshift, clearly illustrating the systematic evolution of the LF with cosmic time. 
At fixed observed magnitude, the number density of galaxies decreases toward higher redshift, reflecting the combined effects of cosmological dimming, galaxy evolution, and the evolving comoving volume element. 
Across all bands, the faint end of the LF becomes progressively truncated at higher redshift, particularly in the shallower redder filters, where survey depth limits the detectability of intrinsically faint galaxies. The vertical lines in Fig.~\ref{fig:lsst_nd_3x2} indicate the LSST $5\sigma$ point-source limiting depths for different survey durations. We adopt the Year~1 to Year~10 LSST coadded depths from the Rubin Observatory survey strategy v5.0 simulations \citep{peter_yoachim_2025_15832326}.
The solid vertical line corresponds to the first-year coadded depth, dotted lines denote intermediate coadds from years two through nine, and the dashed vertical line marks the full ten-year coadded depth. 
The exact numerical values of these median coadded depths are listed in Table~\ref{tab:lsst_coadd_depth}.

\begin{figure}[ht]
\centering
\includegraphics[width=0.4\textwidth]{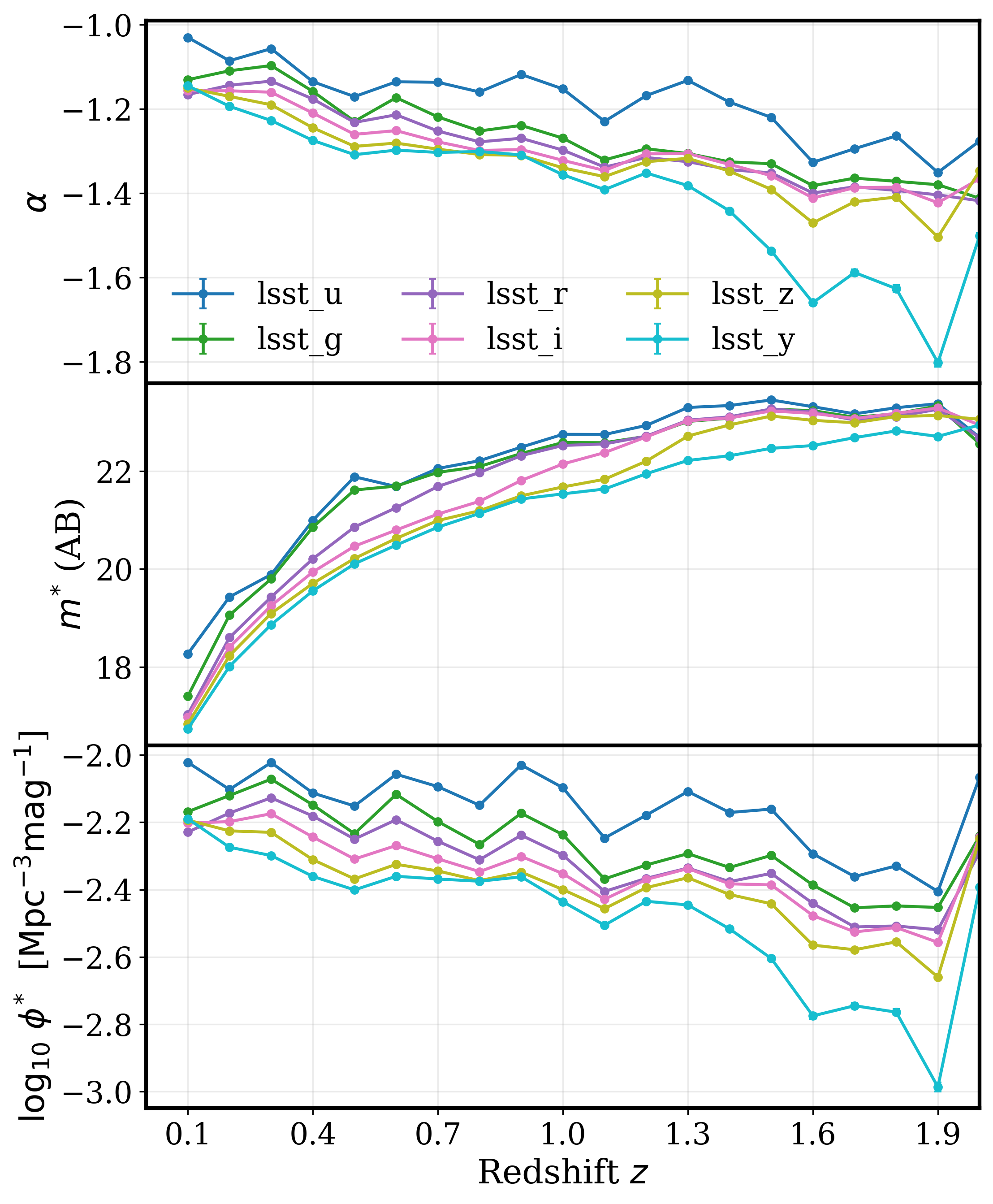}
\caption{Best-fit Schechter function parameters for apparent-magnitude luminosity functions in LSST bands at different redshifts. Uncertainties represent $1\sigma$ errors derived from Poisson statistics of the galaxy counts in each magnitude bin.}
    
\label{fig:schechter}
\end{figure}

As the survey depth increases with time, the accessible faint-magnitude regime expands significantly, leading to a substantial increase in the number of detectable galaxies at all redshifts. This effect is most pronounced in the $g$, $r$, and $i$ bands, where LSST achieves its deepest sensitivities, enabling robust LF measurements over a wide magnitude range even at intermediate redshifts.

To provide an analytical description of these luminosity functions, we model each LF using the standard Schechter form in magnitude space,
\begin{equation}
\phi(m)=0.4\ln 10 \; \phi^{*} \;
10^{0.4(\alpha+1)(m^{*}-m)}
\exp\!\left[-10^{0.4(m^{*}-m)}\right],
\label{eq:schechter_m}
\end{equation}
where $\phi^{*}$ is the normalization, $m^{*}$ is the characteristic magnitude marking the transition between the faint-end power law and the bright-end exponential cutoff, and $\alpha$ is the faint-end slope. 
The binned LF is constructed from galaxy counts in the simulation volume as
\begin{equation}
\phi(m_i)=\frac{N_i}{V\,\Delta m},
\end{equation}
with $N_i$ the number of galaxies in bin $i$, $\Delta m$ the bin width, and $V$ the comoving simulation volume. 
Using the best-fit parameters together with Eq.~(\ref{eq:schechter_m}), the full LF can be reconstructed analytically at each redshift.

Figure~\ref{fig:schechter} shows the redshift evolution of the best-fit Schechter parameters in all LSST bands. 
We find that the faint-end slope $\alpha$ decreases with increasing redshift, becoming more negative toward $z \sim 2$. 
A more negative $\alpha$ corresponds to a steeper faint-end slope and therefore a relatively larger contribution from faint galaxies at higher redshift. 
The characteristic magnitude $m^{*}$ systematically increases with redshift, indicating that the knee of the luminosity function shifts toward fainter magnitudes at earlier cosmic times. 
Meanwhile, the normalization parameter $\phi^{*}$ decreases toward higher redshift, reflecting the declining overall number density of galaxies. 
Together, these trends capture the combined evolution of galaxy luminosity and abundance across cosmic time, and the Schechter parameterization provides a compact and practical representation of the LSST luminosity functions over $0<z\leq 2$. The error bars also represent the formal fitting uncertainties derived using Poisson errors in the luminosity function bins and do not include additional contributions from cosmic variance or systematic modeling uncertainties.

\begin{figure*}[ht]
    \centering
    \includegraphics[width=\textwidth]{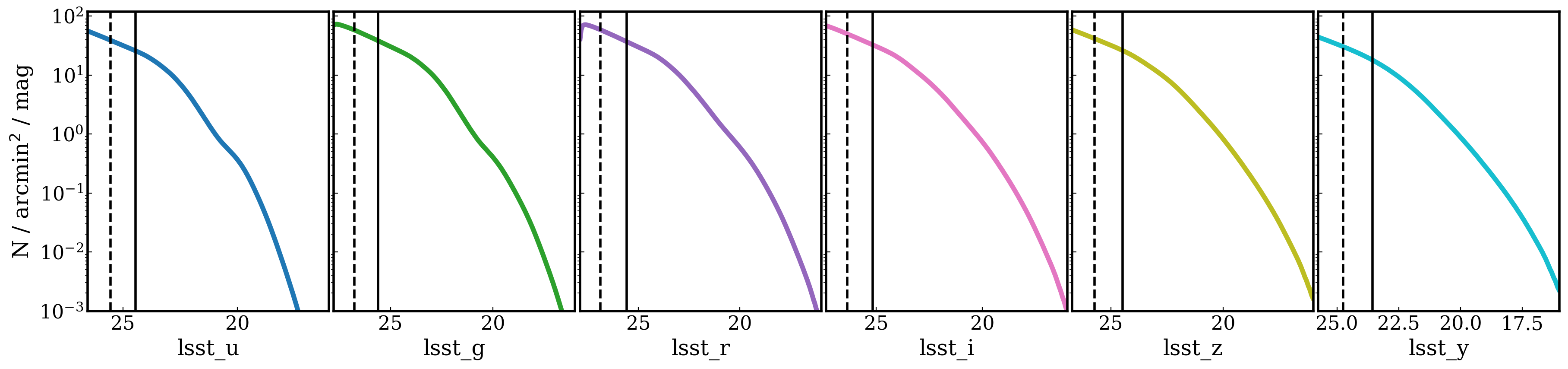}
    \hfill
    \includegraphics[width=0.4\textwidth]{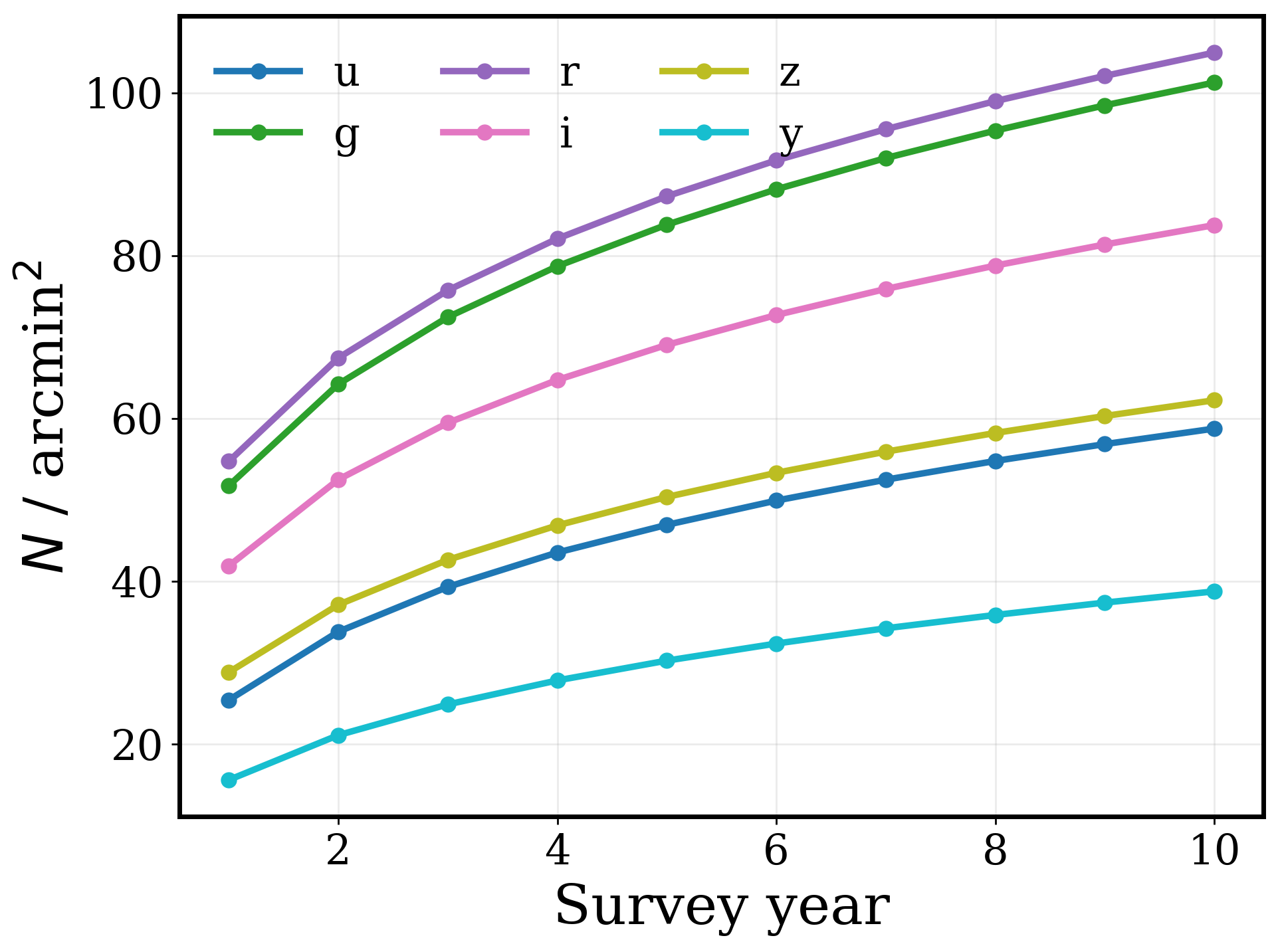}
    \includegraphics[width=0.4\textwidth]{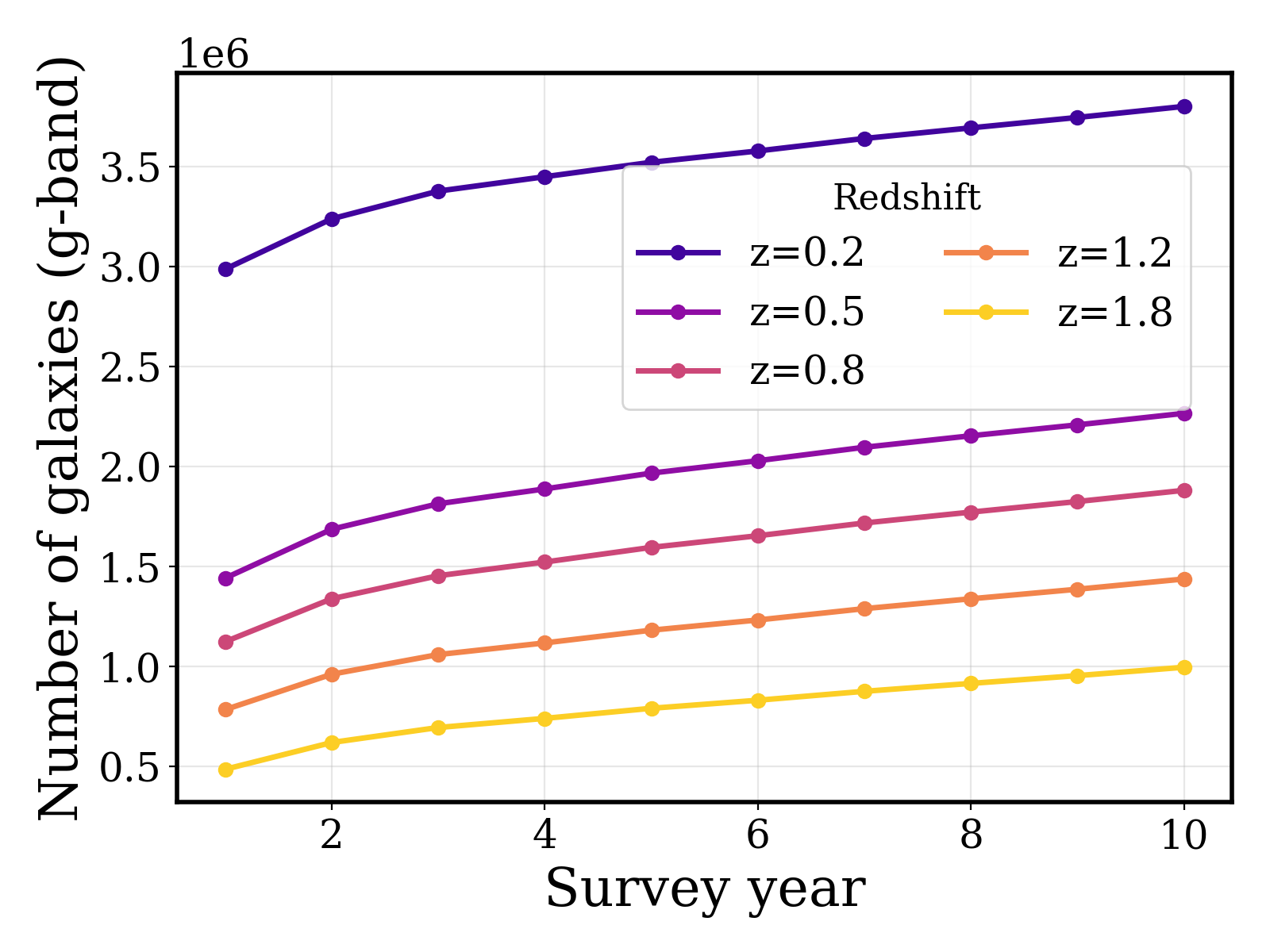}
    \caption{
    Top panel: Differential galaxy number counts, $N(m)$, in the LSST $u$, $g$, $r$, $i$, $z$, and $y$ bands for $0 \leq z \leq 2$, in units of galaxies arcmin$^{-2}$\,mag$^{-1}$. Vertical dashed lines indicate the LSST Year~1 and Year~10  $5\sigma$ depths. 
Bottom panels: Cumulative number of detectable galaxies predicted for LSST. 
Left: Integrated galaxy counts per arcmin$^2$ as a function of survey year for $0 < z \leq 2$, computed by integrating the differential counts up to the coadded depth of each survey year for different LSST bands. 
Right: Cumulative counts in the $g$ band versus survey duration for different redshifts.
    }
    \label{fig:lsst_number_counts}
\end{figure*}

\subsubsection{Evolution of LSST Galaxy Number Density with Survey Depth}

Figure~\ref{fig:lsst_number_counts} summarizes how the progressive increase in LSST coadded depth translates into both differential and cumulative galaxy counts. 
The exact numerical values of the Year~1 through Year~10 median coadded $5\sigma$ depths used in this analysis are listed in Table~\ref{tab:lsst_coadd_depth}.

For each simulated galaxy, photometric magnitude errors in the six LSST bands at LSST Year~1 to Year~10 are computed using the           
  \texttt{photerr} package \citep{crenshaw2024}, which implements the signal-to-noise model of \citet{Ivezic2019}. We treat each 
  galaxy as an extended source using the Gaussian Aperture and PSF (GAAP) model, with the stellar half-mass radius,          
   converted from physical to angular radius using the Planck 2015 cosmology \citep{planck2015}, adopted as both the semi-major and semi-minor axis. The assumed PSF full-width at half-maximum at zenith is ${u{:}1.22, g{:}1.09, r{:}1.02, i{:}0.99, z{:}1.01, y{:}0.99}$ arcsec. The extended magnitude errors, shown in Table~\ref{tab:lsst_coadd_depth}, are computed by taking the average of galaxies with signal-to-noise ratios equal to $5$, or magnitude errors at $\sim 0.217$. The magnitude errors are attached in the data products. 

The top panel of Figure~\ref{fig:lsst_number_counts} shows the differential number counts, $N(m)$, integrated over the full redshift range $0 < z \leq 2$, in the six LSST bands. 
The vertical dashed lines indicate the Year~1 and Year~10 limiting magnitudes for the extended objects (see Table~\ref{tab:lsst_coadd_depth} for exact values). 
As the limiting magnitude shifts to fainter values over the ten-year survey, the accessible faint-end regime expands significantly. 
The differential counts rise steeply toward faint magnitudes before turning over due to the intrinsic decline of the luminosity function and survey incompleteness. 
Among the filters, the $r$ and $i$ bands reach the largest counts, consistent with their deeper coadded sensitivities, while the $u$ and especially $y$ bands show comparatively lower counts due to shallower depth and wavelength-dependent SED effects.

The bottom-left panel of Figure~\ref{fig:lsst_number_counts} shows the integrated number of detectable galaxies per arcmin$^{2}$ as a function of survey year for each LSST band, computed over the redshift range $0<z\le2$. 
The counts are obtained by integrating the differential number counts up to the coadded depth corresponding to each survey year, while applying a photometric uncertainty selection such that only galaxies with extended magnitude errors $ \le 0.22$ are included. 
This threshold effectively selects galaxies with sufficiently well-measured photometry.
In all filters, the number of detectable galaxies increases monotonically with survey duration as the survey depth improves. 
The growth is most pronounced during the early years, when the coadded depth increases most rapidly (Table~\ref{tab:lsst_coadd_depth}), and gradually slows as the survey approaches its ten-year limit. 
The relative ordering of the bands follows the depth hierarchy: the $r$ and $g$ bands yield the highest galaxy densities, while the $u$ and $y$ bands produce fewer detections due to their shallower effective sensitivities.

The bottom-right panel isolates the redshift dependence by showing cumulative counts in the $g$ band for representative redshifts ($z=0.2$, $0.5$, $0.8$, $1.2$, and $1.8$). 
Lower redshift bins contribute substantially more galaxies at every survey year, while higher redshift bins remain comparatively sparse. 
This reflects the combined effects of cosmological dimming, luminosity evolution, and the reduced detectability of intrinsically faint galaxies at larger distances. 
Although increasing survey depth boosts counts at all redshifts, the dominance of low-redshift galaxies persists throughout the ten-year survey.

Overall, these results demonstrate how the gradual extension of the limiting magnitude (Table~\ref{tab:lsst_coadd_depth}) directly drives the increase in detectable galaxy populations, expanding the faint-magnitude regime while maintaining a redshift distribution strongly weighted toward lower-$z$ galaxies.

\begin{figure}
    \centering
    \includegraphics[width=0.45\textwidth]{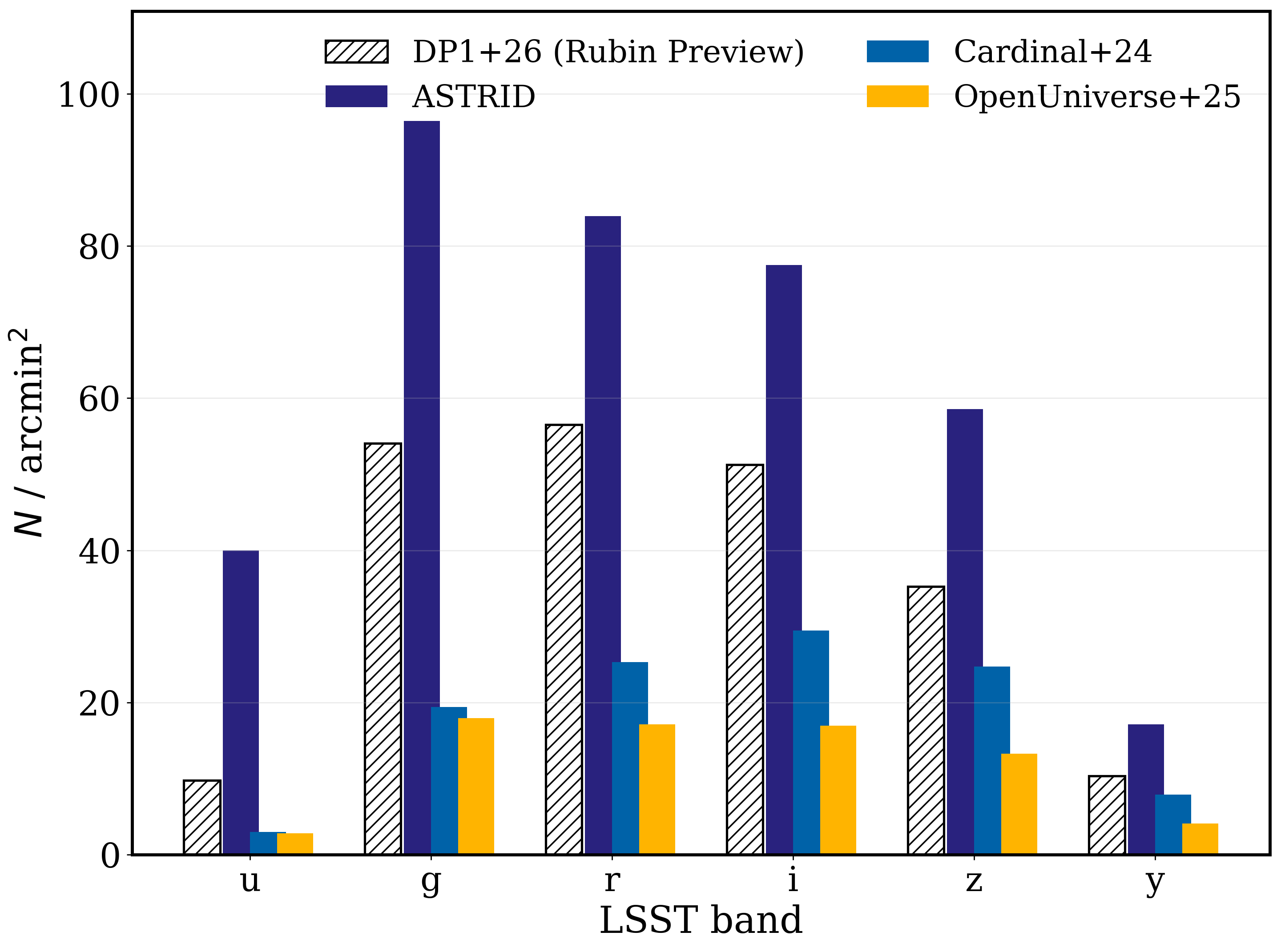}
    \caption{
    Comparison of galaxy number densities in the LSST $u$, $g$, $r$, $i$, $z$, and $y$ bands for ASTRID, Cardinal \citep{to2024buzzard}, OpenUniverse2024 \citep{openuniverse2025}, and Vera C. Rubin Observatory Data Preview 1 \citep{team2026vera}.
    }
    \label{fig:number_density}
\end{figure}

To further place our predictions in the context of existing survey scale mock catalogs, Figure~\ref{fig:number_density} compares the predicted LSST galaxy surface densities from ASTRID with those from the Cardinal \citep{to2024buzzard} and OpenUniverse2024 \citep{openuniverse2025} mock catalogs, together with the Rubin DP1 preview data \citep{team2026vera}. Cardinal is based on empirically calibrated galaxy halo connection models applied to large dark matter-only simulations, while OpenUniverse2024 combines synthetic sky generation with semi analytic and observationally calibrated galaxy population models. To enable a direct comparison between all catalogs, we apply the same DP1 magnitude limits in all LSST bands: $u = 24.55$, $g = 26.18$, $r = 25.96$, $i = 25.71$, $z = 25.07$, and $y = 23.1$. In the $r$ band, for example, ASTRID predicts a galaxy surface density of $\sim84$ arcmin$^{-2}$, compared to $\sim57$ arcmin$^{-2}$ in the Rubin DP1 preview data, $\sim25$ arcmin$^{-2}$ in Cardinal, and $\sim17$ arcmin$^{-2}$ in OpenUniverse2024. Similar trends are observed in the $g$ and $i$ bands, where ASTRID systematically predicts larger galaxy counts than the empirical and semi-analytic mock catalogs while remaining broadly consistent with the overall band-dependent trends seen in DP1. These differences likely reflect variations in the underlying galaxy formation prescriptions, luminosity modeling, dust attenuation treatments, and survey selection assumptions adopted by the different mock frameworks.

\subsubsection{$g-r$ Color Evolution with Redshift}

\begin{figure}
    \centering
    \includegraphics[width=0.45\textwidth]{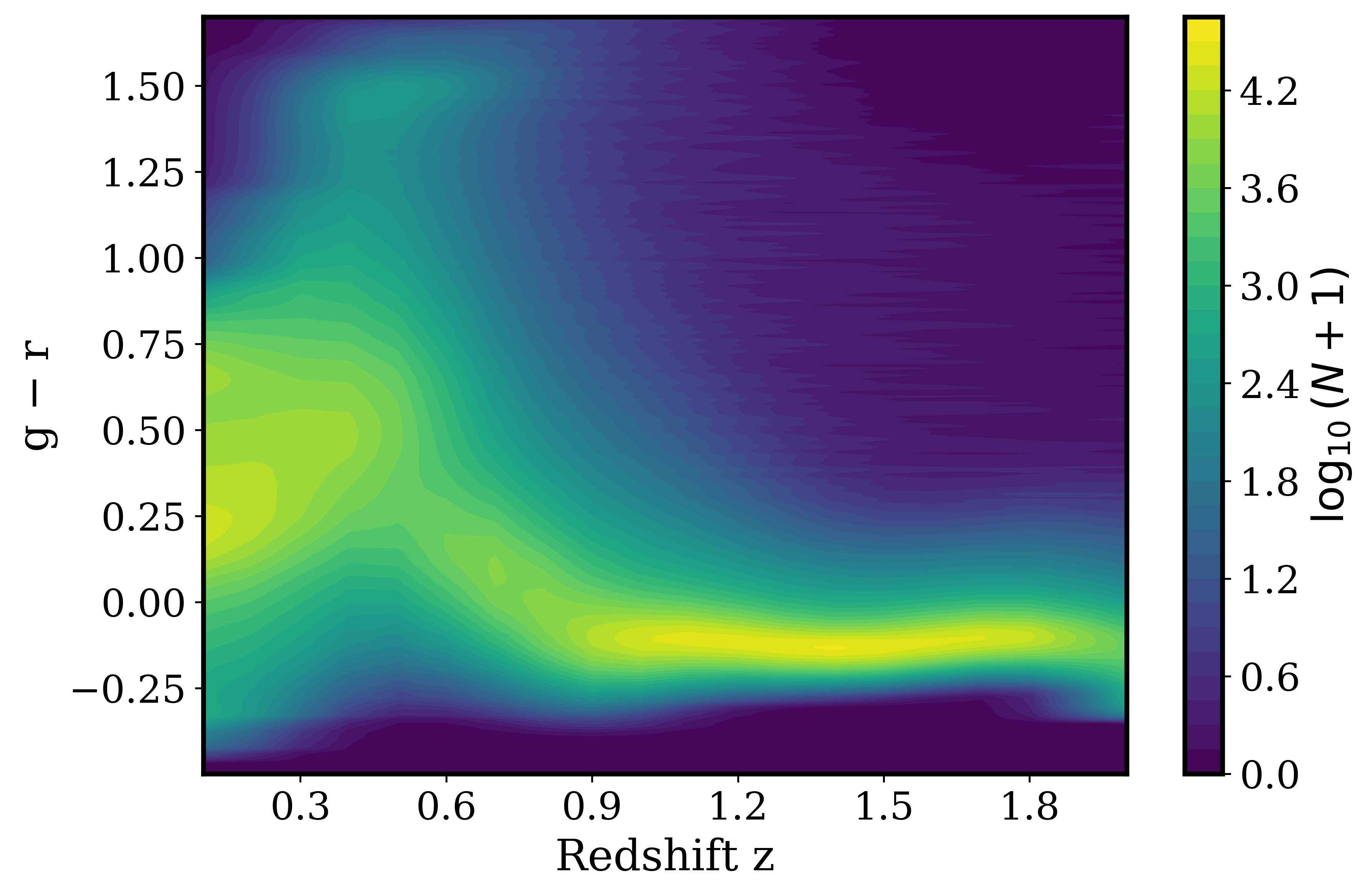}
    \caption{
    Observed $g-r$ color distribution as a function of redshift for LSST Year~10 depth. 
Contours show $\log_{10}(N+1)$ from the ASTRID mock catalog, applying the cuts $m_g < 26.76$ and $m_r < 26.88$. 
    }
    \label{fig:gr_color_z}
\end{figure}

Figure~\ref{fig:gr_color_z} shows the distribution of galaxy $g-r$ colors as a function of redshift, where the color scale represents $\log_{10}(N+1)$. 
At low redshift ($z \lesssim 0.5$), a clear bimodal structure is visible. 
Two distinct sequences emerge: a red population centered around $g-r \sim 0.6$--$0.8$, corresponding to quenched, passive galaxies, and a bluer population around $g-r \sim 0.2$--$0.4$, corresponding to actively star-forming systems. 
This bimodality is consistent with the well-established separation between the red sequence and the blue cloud observed in local galaxy surveys \citep[e.g.,][]{strateva2001color,blanton2003broadband, Baldry2004}. 
The existence of this bimodal structure reflects differences in star formation activity, stellar age, and dust attenuation among galaxy populations.

As redshift increases beyond $z \sim 0.5$, the color distribution gradually shifts toward bluer values, and the red sequence becomes less pronounced. 
By $z \gtrsim 1$, the dominant population is concentrated near $g-r \sim 0$ or slightly negative values, indicating younger stellar populations and enhanced star formation activity at earlier cosmic times. 
This trend is consistent with the cosmic star formation history, which peaks around $z \sim 1$--2 \citep[e.g.,][]{Madau2014}, leading to galaxies being, on average, younger and therefore intrinsically bluer.

In addition to stellar population evolution, observational effects also contribute to this shift. 
At higher redshifts, the observed $g$ and $r$ bands probe progressively bluer rest-frame wavelengths due to $k$-correction, effectively sampling more UV-dominated emission from young stars. 
This naturally drives observed $g-r$ colors toward bluer values with increasing redshift \citep[e.g.,][]{Hogg2002}. 
Furthermore, the relative abundance of massive, quenched galaxies decreases toward higher redshift \citep[e.g.,][]{Ilbert2013}, reducing the prominence of the red sequence.

Overall, the evolution seen in Figure~\ref{fig:gr_color_z} reflects the combined impact of stellar population aging, declining quenching efficiency at earlier times, cosmological band shifting, and the evolving mass function of galaxies. 
The gradual disappearance of the strong bimodality beyond $z \sim 0.5$ is therefore consistent with both observational surveys and theoretical models of galaxy evolution.

\subsection{Color--Morphology Relation at $z=0$}

Using galaxy color and magnitude information provides a practical way to connect observable galaxy properties to their underlying physical state and is increasingly important for mitigating astrophysical systematics such as intrinsic alignments in large-scale structure surveys \citep{Krause_2015,10.1093/mnras/stz2197,Secco_2022,2024arXiv241022272M}.

Hydrodynamical simulations provide an essential testbed for studying these relationships and connecting observable galaxy properties to their physical origins \citep{Nelson_2017,10.1093/mnras/stv2154,Tenneti_2016,Samuroff_2021,2023MNRAS.523.5899D}. Since reliable shape measurements require a sufficiently large number of stellar particles per galaxy, we restrict our analysis to galaxies with stellar masses above $M_* \geq 10^{9.5} M_\odot$.
\begin{figure}
    \centering
    \includegraphics[width=0.45\textwidth]{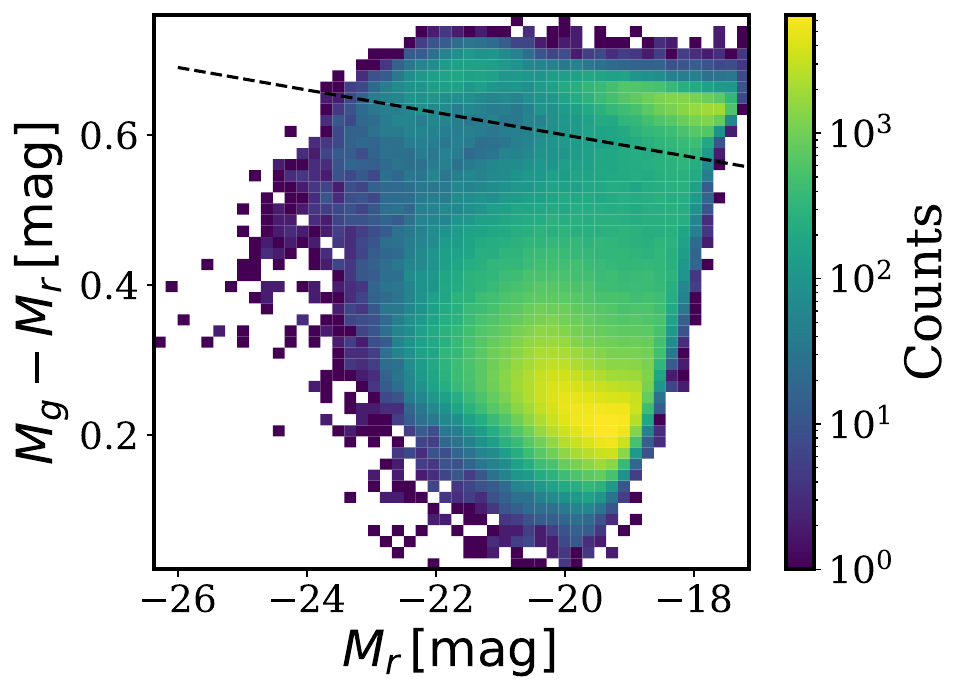}
    \caption{Color-Magnitude histogram of Astrid galaxies above $M_* \geq 10^{9.5} M_\odot$ with the simple red/blue cut of this analysis as a black dashed line. }
    \label{fig:color_mag}
\end{figure}

To test whether ASTRID reproduces the expected CMR behavior for LSST, we apply a color--magnitude split in $M_g-M_r$ and $M_r$ space using the relation defined in Section~\ref{subsec:morphology}. 
This selection yields approximately $5.2\times10^4$ red galaxies and $5.0\times10^5$ blue galaxies (Figure~\ref{fig:color_mag}).

\begin{figure}
    \centering
    \includegraphics[width=0.45\textwidth]{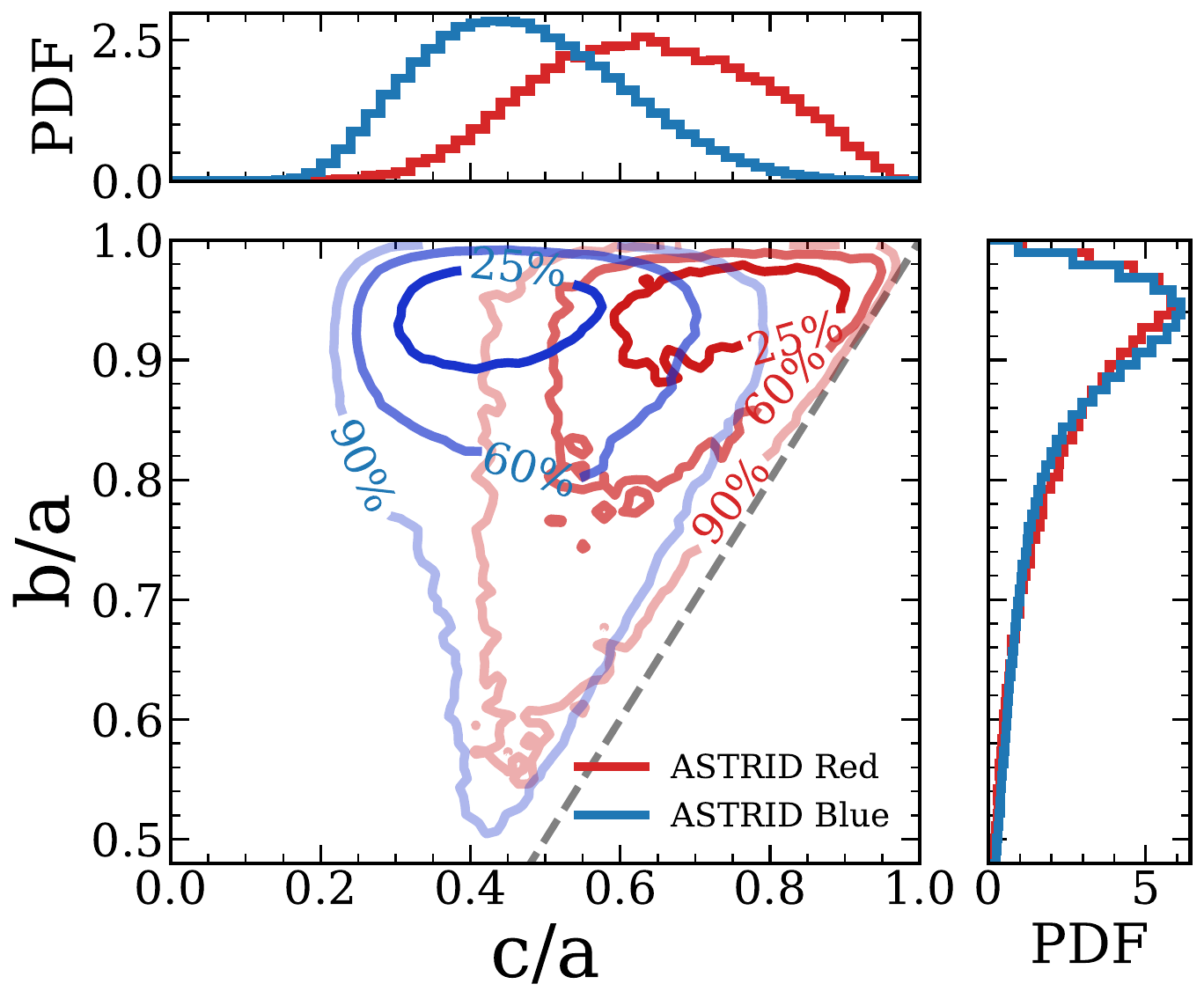}
    \caption{2D contours depicting the distribution of the measured axis ratios for red and blue ASTRID galaxies using LSST colors and their 1D PDFs. The number of galaxies in our selection is shown in the bottom-left corner, and the enclosed fractions are shown along the contour lines.}
    \label{fig:axis_ratios}
\end{figure}

\begin{figure}[ht]
    \centering
    \includegraphics[width=0.45\textwidth]{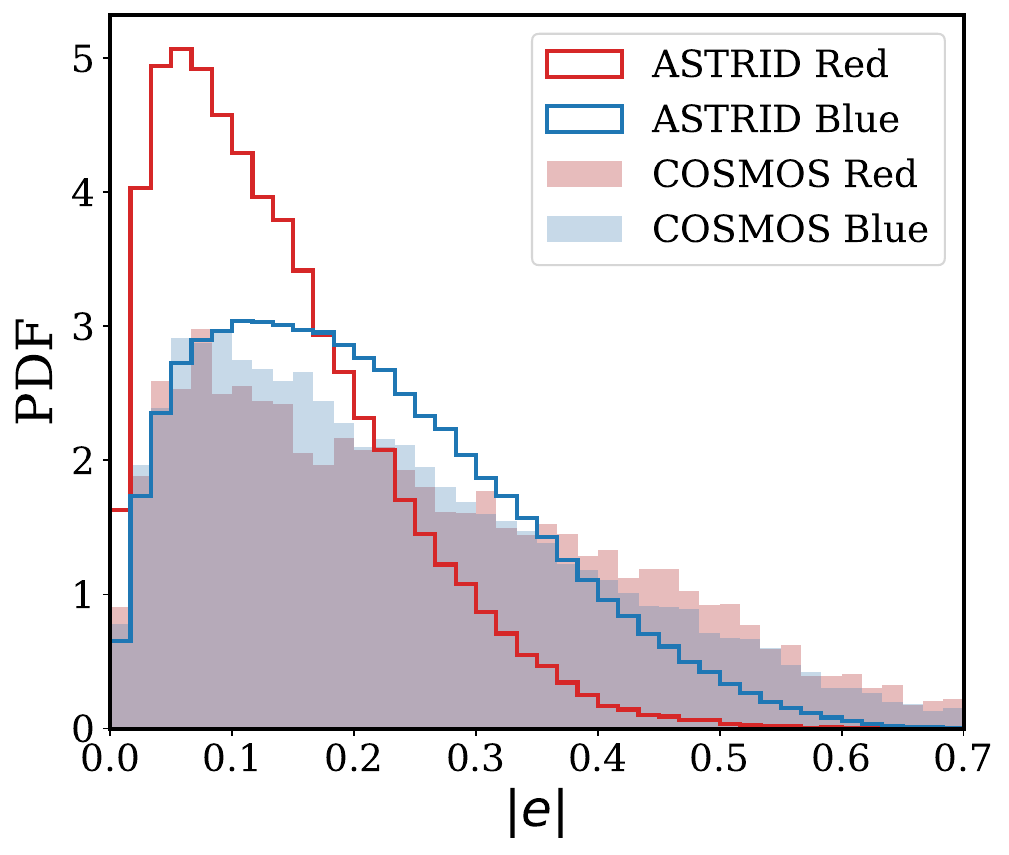}
    \caption{Probability distribution of projected ellipticities of the ASTRID red and blue samples using LSST colors alongside COSMOS2020 measured ellipticities using similar color-magnitude and stellar-mass cuts.  \cite{2022ApJS..258...11W}.}
    \label{fig:cosmos}
\end{figure}

We first assess the distribution of galaxy shapes using the principal axis ratios derived from the inertia tensor (Figure~\ref{fig:axis_ratios}). Given the axis ratio $b/a$ is close to 1 for all samples, a small $c/a$ indicates that the galaxy shapes are oblate, whereas a larger $c/a$ indicates near-spherical shapes. We find that the general population of red galaxies in our sample is more spherical relative to that of blue galaxies, with average $c/a$ axis ratios of 0.64 and 0.47, respectively. The results for the blue distribution depend slightly on the definition of the galaxy center, where we take it to be the minimum potential in this work.
We also compare the projected ellipticity distributions of the simulated galaxies to observational measurements from the COSMOS survey \citep{2022ApJS..258...11W} (Figure \ref{fig:cosmos}). To enable a consistent comparison, we utilize the Classic COSMOS2020 catalog to retrieve galaxy shape,  stellar-mass estimates, and Subaru rest-frame magnitudes for galaxies up to redshift $z\leq 1$ to apply a similar classification used in this work to separate red and blue galaxy populations for both the ASTRID and COSMOS samples. We find that the projected ellipticity distributions are in overall good agreement between the simulation and observations. For the red population, both the COSMOS and ASTRID distributions peak near $|e|\sim0.07$. For the blue galaxy population, the COSMOS distribution peaks near $|e|\sim0.09$, while the ASTRID distribution peaks at $|e|\sim0.11$, which remains reasonably close given the simplicity of the classification and shape measurement methods.

\section{Discussion}
\label{sec:discussion}
In this work, we constructed and validated forward-modelled galaxy photometric catalogs for LSST using the \textsc{ASTRID} cosmological hydrodynamical simulation. We computed intrinsic galaxy luminosities by summing \textsc{FSPS}-based stellar population synthesis predictions over star particles treated as simple stellar populations, and we implemented a physically motivated dust attenuation model in which the optical depth scales with metal surface density and follows a power-law wavelength dependence. Calibrating the dust parameters at $z=0$ using SDSS luminosity functions and validating the same model at $z=0.5$, $1.0$, and $1.5$ in rest-frame $BVRI$, we found that the attenuated ASTRID luminosity functions reproduce existing observational constraints across multiple bands and redshifts. Leveraging this validated framework, we produced LSST-ready mock photometric catalogs over $0 \le z \le 2$ (in steps of $\Delta z=0.1$), and we presented predictions for apparent-magnitude luminosity functions in the LSST $ugrizy$ bands, their Schechter-function representations, and the implied differential and cumulative galaxy number counts as a function of survey depth from Year~1 to Year~10. We also performed a preliminary analysis of galaxy morphology and its connection to color, finding that the simulation generally reproduces the color--morphology relation (CMR) at $z=0$ expected of hydrodynamical simulations. A more detailed study of galaxy shapes and intrinsic alignments will be presented in future work.

A key strength of our approach is that the calibrated dust treatment enables a good match to observed luminosity functions while preserving predictive power for LSST depths, where surveys will access substantially fainter galaxies than many existing LF measurements. Because ASTRID combines a large cosmological volume with high mass resolution, our mock catalogs naturally extend into the faint-end regime that will be critical for LSST science, including number-count forecasting and selection-function studies. Interestingly, the galaxy number densities predicted by ASTRID for LSST depths are generally closer to the current LSST DP1 \citep{team2026vera} measurements than those inferred from other mock catalogs such as Cardinal \citep{to2024buzzard} and OpenUniverse \citep{openuniverse2025}. In addition, our LSST forward modeling reveals clear structure in the evolution of galaxy colors with redshift. In particular, we find a prominent bimodality in $g-r$ at low redshift and an overall trend in which the typical $g-r$ color becomes redder up to $z\sim0.5$ and then shifts toward bluer values at higher redshift, consistent with the combined effects of evolving stellar populations and bandpass shifting with redshift. To support downstream applications, we also make available the intrinsic and dust-attenuated magnitudes for both galaxies and their constituent star particles, enabling flexible reanalysis (e.g., alternative apertures, selection cuts, and comparisons to future LSST data products).

Despite these successes, our predictions inherit several limitations that motivate future improvements. First, our intrinsic luminosities rely on a single SPS framework and a grid-based interpolation scheme; while computationally efficient, it does not capture the full diversity of SPS systematics (e.g., treatment of short-lived stellar phases) and does not include radiative transfer effects. Second, our dust attenuation prescription is intentionally simple (two-parameter, metal-surface-density-based, power-law wavelength scaling) and does not explicitly incorporate geometry, scattering, or age-dependent attenuation, which can be important for detailed color distributions and for interpreting galaxy types across cosmic time. Future work could therefore compare against more sophisticated dust treatments (e.g., radiative-transfer post-processing such as \textsc{SKIRT}) and explore extensions such as two-component (birth-cloud + diffuse ISM) or inclination-dependent attenuation models. Future work will also quantify the role of AGN feedback in shaping observable galaxy properties and cosmological statistics within LSST mock catalogs, including its impact on galaxy colors, number densities, intrinsic alignments, and the matter power spectrum. Finally, once early LSST data become available, the framework developed here can be revalidated directly in observed-frame LSST photometry, enabling more stringent constraints on dust and stellar population modeling and more precise forecasts for LSST number counts and color--redshift distributions.
\section*{Data Availability}
The ASTRID simulations are publicly available through \url{https://astrid.psc.edu}, where we provide access to the full particle-in-group data, the FOFGroup and SubFind subgroup catalogs, the MBH merger catalog, as well as the LSST mock observation products generated from the simulations.

\section*{Acknowledgements}
ASTRID is part of the Frontera computing project at the Texas Advanced Computing Center. Frontera is made possible by NSF award OAC-1818253. TZ acknowledges funding from Schmidt Sciences. YZ and TDM acknowledge the support from the NASA FINESST grant NNH24ZDA001N. 
TDM acknowledges funding from NASA ATP 80NSSC20K0519, NSF PHY-2020295, NASA ATP NNX17AK56G, and NASA ATP 80NSSC18K101, NASA Theory grant 80NSSC22K072. SB acknowledges funding from NASA ATP 80NSSC22K1897 and NSF AST-2509639.

\bibliography{ref}

@article{driver2012galaxy,
  title={Galaxy And Mass Assembly (GAMA): the 0.013< z< 0.1 cosmic spectral energy distribution from 0.1 $\mu$m to 1 mm},
  author={Driver, Simon P and Robotham, Aaron SG and Kelvin, L and Alpaslan, Mehmet and Baldry, Ivan K and Bamford, Steven P and Brough, Sarah and Brown, Michael and Hopkins, Andrew M and Liske, Jochen and others},
  journal={Monthly Notices of the Royal Astronomical Society},
  volume={427},
  number={4},
  pages={3244--3264},
  year={2012},
  publisher={Blackwell Science Ltd Oxford, UK}
}

@article{loveday2012galaxy,
  title={Galaxy and Mass Assembly (GAMA): ugriz galaxy luminosity functions},
  author={Loveday, Jon and Norberg, Peder and Baldry, Ivan K and Driver, Simon P and Hopkins, Andrew M and Peacock, John A and Bamford, Steven P and Liske, Jochen and Bland-Hawthorn, Joss and Brough, Sarah and others},
  journal={Monthly Notices of the Royal Astronomical Society},
  volume={420},
  number={2},
  pages={1239--1262},
  year={2012},
  publisher={Blackwell Publishing Ltd Oxford, UK}
}

@ARTICLE{Pei1992,
       author = {{Pei}, Yichuan C.},
        title = "{Interstellar Dust from the Milky Way to the Magellanic Clouds}",
      journal = {\apj},
         year = 1992,
        month = aug,
       volume = {395},
        pages = {130},
          doi = {10.1086/171637},
       adsurl = {https://ui.adsabs.harvard.edu/abs/1992ApJ...395..130P}
}

@ARTICLE{lachance2024,
       author = {{LaChance}, Patrick and {Croft}, Rupert and {Ni}, Yueying and {Chen}, Nianyi and {Matteo}, Tiziana Di and {Bird}, Simeon},
        title = "{The evolution of galaxy morphology from redshift z=6 to 3: Mock JWST observations of galaxies in the ASTRID simulation}",
      journal = {The Open Journal of Astrophysics},
         year = 2025,
        month = feb,
       volume = {8},
          eid = {20},
        pages = {20},
          doi = {10.33232/001c.129991},
archivePrefix = {arXiv},
       eprint = {2401.16608},
 primaryClass = {astro-ph.GA},
       adsurl = {https://ui.adsabs.harvard.edu/abs/2025OJAp....8E..20L}
}

@ARTICLE{Wilkins2017,
       author = {{Wilkins}, Stephen M. and {Feng}, Yu and {Di Matteo}, Tiziana and {Croft}, Rupert and {Lovell}, Christopher C. and {Waters}, Dacen},
        title = "{The properties of the first galaxies in the BlueTides simulation}",
      journal = {\mnras},
         year = 2017,
        month = aug,
       volume = {469},
       number = {3},
        pages = {2517-2530},
          doi = {10.1093/mnras/stx841},
archivePrefix = {arXiv},
       eprint = {1704.00954},
 primaryClass = {astro-ph.GA},
       adsurl = {https://ui.adsabs.harvard.edu/abs/2017MNRAS.469.2517W}
}

@ARTICLE{Calzetti2000,
       author = {{Calzetti}, Daniela and {Armus}, Lee and {Bohlin}, Ralph C. and {Kinney}, Anne L. and {Koornneef}, Jan and {Storchi-Bergmann}, Thaisa},
        title = "{The Dust Content and Opacity of Actively Star-forming Galaxies}",
      journal = {\apj},
         year = 2000,
        month = apr,
       volume = {533},
       number = {2},
        pages = {682-695},
          doi = {10.1086/308692},
archivePrefix = {arXiv},
       eprint = {astro-ph/9911459},
 primaryClass = {astro-ph},
       adsurl = {https://ui.adsabs.harvard.edu/abs/2000ApJ...533..682C}
}

@ARTICLE{Chabrier2003,
       author = {{Chabrier}, Gilles},
        title = "{Galactic Stellar and Substellar Initial Mass Function}",
      journal = {\pasp},
         year = 2003,
        month = jul,
       volume = {115},
       number = {809},
        pages = {763-795},
          doi = {10.1086/376392},
archivePrefix = {arXiv},
       eprint = {astro-ph/0304382},
 primaryClass = {astro-ph},
       adsurl = {https://ui.adsabs.harvard.edu/abs/2003PASP..115..763C}
}

@ARTICLE{Sanchez-Blazquez2006_miles,
       author = {{S{\'a}nchez-Bl{\'a}zquez}, P. and {Peletier}, R.~F. and {Jim{\'e}nez-Vicente}, J. and {Cardiel}, N. and {Cenarro}, A.~J. and {Falc{\'o}n-Barroso}, J. and {Gorgas}, J. and {Selam}, S. and {Vazdekis}, A.},
        title = "{Medium-resolution Isaac Newton Telescope library of empirical spectra}",
      journal = {\mnras},
         year = 2006,
        month = sep,
       volume = {371},
       number = {2},
        pages = {703-718},
          doi = {10.1111/j.1365-2966.2006.10699.x},
archivePrefix = {arXiv},
       eprint = {astro-ph/0607009},
 primaryClass = {astro-ph},
       adsurl = {https://ui.adsabs.harvard.edu/abs/2006MNRAS.371..703S}
}

@ARTICLE{Bressan2012_parsec,
       author = {{Bressan}, Alessandro and {Marigo}, Paola and {Girardi}, L{\'e}o. and {Salasnich}, Bernardo and {Dal Cero}, Claudia and {Rubele}, Stefano and {Nanni}, Ambra},
        title = "{PARSEC: stellar tracks and isochrones with the PAdova and TRieste Stellar Evolution Code}",
      journal = {\mnras},
         year = 2012,
        month = nov,
       volume = {427},
       number = {1},
        pages = {127-145},
          doi = {10.1111/j.1365-2966.2012.21948.x},
archivePrefix = {arXiv},
       eprint = {1208.4498},
 primaryClass = {astro-ph.SR},
       adsurl = {https://ui.adsabs.harvard.edu/abs/2012MNRAS.427..127B}
}

@ARTICLE{Conroy2010,
       author = {{Conroy}, Charlie and {Gunn}, James E.},
        title = "{The Propagation of Uncertainties in Stellar Population Synthesis Modeling. III. Model Calibration, Comparison, and Evaluation}",
      journal = {\apj},
         year = 2010,
        month = apr,
       volume = {712},
       number = {2},
        pages = {833-857},
          doi = {10.1088/0004-637X/712/2/833},
archivePrefix = {arXiv},
       eprint = {0911.3151},
 primaryClass = {astro-ph.CO},
       adsurl = {https://ui.adsabs.harvard.edu/abs/2010ApJ...712..833C}
}

@ARTICLE{Conroy2009,
       author = {{Conroy}, Charlie and {Gunn}, James E. and {White}, Martin},
        title = "{The Propagation of Uncertainties in Stellar Population Synthesis Modeling. I. The Relevance of Uncertain Aspects of Stellar Evolution and the Initial Mass Function to the Derived Physical Properties of Galaxies}",
      journal = {\apj},
         year = 2009,
        month = jul,
       volume = {699},
       number = {1},
        pages = {486-506},
          doi = {10.1088/0004-637X/699/1/486},
archivePrefix = {arXiv},
       eprint = {0809.4261},
 primaryClass = {astro-ph},
       adsurl = {https://ui.adsabs.harvard.edu/abs/2009ApJ...699..486C}
}

@ARTICLE{Kauffmann2003,
       author = {{Kauffmann}, Guinevere and {Heckman}, Timothy M. and {White}, Simon D.~M. and {Charlot}, St{\'e}phane and {Tremonti}, Christy and {Brinchmann}, Jarle and {Bruzual}, Gustavo and {Peng}, Eric W. and {Seibert}, Mark and {Bernardi}, Mariangela and {Blanton}, Michael and {Brinkmann}, Jon and {Castander}, Francisco and {Cs{\'a}bai}, Istvan and {Fukugita}, Masataka and {Ivezic}, Zeljko and {Munn}, Jeffrey A. and {Nichol}, Robert C. and {Padmanabhan}, Nikhil and {Thakar}, Aniruddha R. and {Weinberg}, David H. and {York}, Donald},
        title = "{Stellar masses and star formation histories for {}10$^{5}$ galaxies from the Sloan Digital Sky Survey}",
      journal = {\mnras},
         year = 2003,
        month = may,
       volume = {341},
       number = {1},
        pages = {33-53},
          doi = {10.1046/j.1365-8711.2003.06291.x},
archivePrefix = {arXiv},
       eprint = {astro-ph/0204055},
 primaryClass = {astro-ph},
       adsurl = {https://ui.adsabs.harvard.edu/abs/2003MNRAS.341...33K}
}

@ARTICLE{Baldry2006,
       author = {{Baldry}, I.~K. and {Balogh}, M.~L. and {Bower}, R.~G. and {Glazebrook}, K. and {Nichol}, R.~C. and {Bamford}, S.~P. and {Budavari}, T.},
        title = "{Galaxy bimodality versus stellar mass and environment}",
      journal = {\mnras},
         year = 2006,
        month = dec,
       volume = {373},
       number = {2},
        pages = {469-483},
          doi = {10.1111/j.1365-2966.2006.11081.x},
archivePrefix = {arXiv},
       eprint = {astro-ph/0607648},
 primaryClass = {astro-ph},
       adsurl = {https://ui.adsabs.harvard.edu/abs/2006MNRAS.373..469B}
}

@ARTICLE{Bird2022,
       author = {{Bird}, Simeon and {Ni}, Yueying and {Di Matteo}, Tiziana and {Croft}, Rupert and {Feng}, Yu and {Chen}, Nianyi},
        title = "{The ASTRID simulation: galaxy formation and reionization}",
      journal = {\mnras},
         year = 2022,
        month = may,
       volume = {512},
       number = {3},
        pages = {3703-3716},
          doi = {10.1093/mnras/stac648},
archivePrefix = {arXiv},
       eprint = {2111.01160},
 primaryClass = {astro-ph.GA},
       adsurl = {https://ui.adsabs.harvard.edu/abs/2022MNRAS.512.3703B}
}

@ARTICLE{Ni2022_astrid,
       author = {{Ni}, Yueying and {Di Matteo}, Tiziana and {Bird}, Simeon and {Croft}, Rupert and {Feng}, Yu and {Chen}, Nianyi and {Tremmel}, Michael and {DeGraf}, Colin and {Li}, Yin},
        title = "{The ASTRID simulation: the evolution of supermassive black holes}",
      journal = {\mnras},
         year = 2022,
        month = jun,
       volume = {513},
       number = {1},
        pages = {670-692},
          doi = {10.1093/mnras/stac351},
archivePrefix = {arXiv},
       eprint = {2110.14154},
 primaryClass = {astro-ph.GA},
       adsurl = {https://ui.adsabs.harvard.edu/abs/2022MNRAS.513..670N}
}

@ARTICLE{Ni2024,
       author = {{Ni}, Yueying and {Chen}, Nianyi and {Zhou}, Yihao and {Park}, Minjung and {Yang}, Yanhui and {DiMatteo}, Tiziana and {Bird}, Simeon and {Croft}, Rupert},
        title = "{The Astrid Simulation: Evolution of black holes and galaxies to z=0.5 and different evolution pathways for galaxy quenching}",
      journal = {arXiv e-prints},
         year = 2024,
        month = sep,
          eid = {arXiv:2409.10666},
        pages = {arXiv:2409.10666},
          doi = {10.48550/arXiv.2409.10666},
archivePrefix = {arXiv},
       eprint = {2409.10666},
 primaryClass = {astro-ph.GA},
       adsurl = {https://ui.adsabs.harvard.edu/abs/2024arXiv240910666N}
}

@software{Feng2018,
       author = {{Feng}, Yu and {Bird}, Simeon and {Anderson}, Lauren and {Font-Ribera}, Andreu and {Pedersen}, Chris},
        title = "{MP-Gadget/MP-Gadget: A tag for getting a DOI}",
         year = 2018,
        month = oct,
          eid = {10.5281/zenodo.1451799},
          doi = {10.5281/zenodo.1451799},
      version = {FirstDOI},
    publisher = {Zenodo},
       adsurl = {https://ui.adsabs.harvard.edu/abs/2018zndo...1451799F}
}

@ARTICLE{Pan2023,
       author = {{Pan}, Yue and {Simpson}, Christine M. and {Kravtsov}, Andrey and {G{\'o}mez}, Facundo A. and {Grand}, Robert J.~J. and {Marinacci}, Federico and {Pakmor}, R{\"u}diger and {Manwadkar}, Viraj and {Esmerian}, Clarke J.},
        title = "{Colour and infall time distributions of satellite galaxies in simulated Milky-Way analogues}",
      journal = {\mnras},
         year = 2023,
        month = mar,
       volume = {519},
       number = {3},
        pages = {4499-4513},
          doi = {10.1093/mnras/stac3663},
archivePrefix = {arXiv},
       eprint = {2208.13805},
 primaryClass = {astro-ph.GA},
       adsurl = {https://ui.adsabs.harvard.edu/abs/2023MNRAS.519.4499P}
}

@ARTICLE{Guidi2015,
       author = {{Guidi}, Giovanni and {Scannapieco}, Cecilia and {Walcher}, C. Jakob},
        title = "{Biases and systematics in the observational derivation of galaxy properties: comparing different techniques on synthetic observations of simulated galaxies}",
      journal = {\mnras},
         year = 2015,
        month = dec,
       volume = {454},
       number = {3},
        pages = {2381-2400},
          doi = {10.1093/mnras/stv2050},
archivePrefix = {arXiv},
       eprint = {1507.00347},
 primaryClass = {astro-ph.GA},
       adsurl = {https://ui.adsabs.harvard.edu/abs/2015MNRAS.454.2381G}
}

@article{Baldry2004,
  title={Quantifying the bimodal color-magnitude distribution of galaxies},
  author={Baldry, Ivan K and Glazebrook, Karl and Brinkmann, Jon and Ivezi{\'c}, {\v{Z}}eljko and Lupton, Robert H and Nichol, Robert C and Szalay, Alexander S},
  journal={The Astrophysical Journal},
  volume={600},
  number={2},
  pages={681--694},
  year={2004}
}

@article{strateva2001color,
  title={Color separation of galaxy types in the Sloan Digital Sky Survey imaging data},
  author={Strateva, Iskra and Ivezi{\'c}, {\v{Z}}eljko and Knapp, Gillian R and Narayanan, Vijay K and Strauss, Michael A and Gunn, James E and Lupton, Robert H and Schlegel, David and Bahcall, Neta A and Brinkmann, Jon and others},
  journal={The Astronomical Journal},
  volume={122},
  number={4},
  pages={1861--1874},
  year={2001}
}

@article{blanton2003broadband,
  title={The broadband optical properties of galaxies with redshifts 0.02< z< 0.22},
  author={Blanton, Michael R and Hogg, David W and Bahcall, Neta A and Baldry, Ivan K and Brinkmann, J and Csabai, Istv{\'a}n and Eisenstein, Daniel and Fukugita, Masataka and Gunn, James E and Ivezi{\'c}, {\v{Z}}eljko and others},
  journal={The Astrophysical Journal},
  volume={594},
  number={1},
  pages={186--207},
  year={2003}
}

@article{madau2014,
  title={Cosmic star-formation history},
  author={Madau, Piero and Dickinson, Mark},
  journal={Annual Review of Astronomy and Astrophysics},
  volume={52},
  number={1},
  pages={415--486},
  year={2014},
  publisher={Annual Reviews}
}

@article{hogg2002,
  title={The K correction},
  author={Hogg, David W and Baldry, Ivan K and Blanton, Michael R and Eisenstein, Daniel J},
  journal={arXiv preprint astro-ph/0210394},
  year={2002}
}

@article{ilbert2013,
  title={Mass assembly in quiescent and star-forming galaxies since z≃ 4 from UltraVISTA},
  author={Ilbert, Olivier and McCracken, Henry J and Le F{\`e}vre, Olivier and Capak, Peter and Dunlop, James and Karim, Alexander and Renzini, MA and Caputi, Karina and Boissier, Samuel and Arnouts, St{\'e}phane and others},
  journal={Astronomy \& Astrophysics},
  volume={556},
  pages={A55},
  year={2013},
  publisher={EDP Sciences}
}

@article{pillepich2018,
  title={Simulating galaxy formation with the IllustrisTNG model},
  author={Pillepich, Annalisa and Springel, Volker and Nelson, Dylan and Genel, Shy and Naiman, Jill and Pakmor, R{\"u}diger and Hernquist, Lars and Torrey, Paul and Vogelsberger, Mark and Weinberger, Rainer and others},
  journal={Monthly Notices of the Royal Astronomical Society},
  volume={473},
  number={3},
  pages={4077--4106},
  year={2018},
  publisher={Oxford University Press}
}

@article{nelson2018,
  title={The IllustrisTNG simulations: public data release},
  author={Nelson, Dylan and Springel, Volker and Pillepich, Annalisa and Rodriguez-Gomez, Vicente and Torrey, Paul and Genel, Shy and Vogelsberger, Mark and Pakmor, Ruediger and Marinacci, Federico and Weinberger, Rainer and others},
  journal={Computational Astrophysics and Cosmology},
  volume={6},
  number={1},
  pages={2},
  year={2019},
  publisher={Springer}
}

@article{schaye2015,
  title={The EAGLE project: simulating the evolution and assembly of galaxies and their environments},
  author={Schaye, Joop and Crain, Robert A and Bower, Richard G and Furlong, Michelle and Schaller, Matthieu and Theuns, Tom and Dalla Vecchia, Claudio and Frenk, Carlos S and McCarthy, IG and Helly, John C and others},
  journal={Monthly Notices of the Royal Astronomical Society},
  volume={446},
  number={1},
  pages={521--554},
  year={2015},
  publisher={Oxford University Press}
}

@article{dave2019,
  title={SIMBA: Cosmological simulations with black hole growth and feedback},
  author={Dav{\'e}, Romeel and Angl{\'e}s-Alc{\'a}zar, Daniel and Narayanan, Desika and Li, Qi and Rafieferantsoa, Mika H and Appleby, Sarah},
  journal={Monthly Notices of the Royal Astronomical Society},
  volume={486},
  number={2},
  pages={2827--2849},
  year={2019},
  publisher={Oxford University Press}
}

@article{ilbert2005,
  title={The VIMOS-VLT deep survey-Evolution of the galaxy luminosity function up to z= 2 in first epoch data},
  author={Ilbert, Olivier and Tresse, L and Zucca, E and Bardelli, S and Arnouts, S and Zamorani, G and Pozzetti, L and Bottini, D and Garilli, B and Le Brun, V and others},
  journal={Astronomy \& Astrophysics},
  volume={439},
  number={3},
  pages={863--876},
  year={2005},
  publisher={EDP Sciences}
}

@article{gabasch2004,
  title={The evolution of the luminosity functions in the FORS Deep Field from low to high redshift-I. The blue bands},
  author={Gabasch, Armin and Bender, Ralf and Seitz, Stella and Hopp, Ulrich and Saglia, Roberto P and Feulner, Georg and Snigula, Jan and Drory, Niv and Appenzeller, Immo and Heidt, Jochen and others},
  journal={Astronomy \& Astrophysics},
  volume={421},
  number={1},
  pages={41--58},
  year={2004},
  publisher={EDP Sciences}
}

@article{dahlen2005,
  title={The Evolution of the Optical and Near-Infrared Galaxy Luminosity Functions and Luminosity Densities to z\~{} 2},
  author={Dahlen, Tomas and Mobasher, Bahram and Somerville, Rachel S and Moustakas, Leonidas A and Dickinson, Mark and Ferguson, Henry C and Giavalisco, Mauro},
  journal={The Astrophysical Journal},
  volume={631},
  number={1},
  pages={126--144},
  year={2005}
}

@article{ramos2011,
  title={Evolution of galaxy luminosity function using photometric redshifts},
  author={Ramos, BHF and Pellegrini, PS and Benoist, C and da Costa, LN and Maia, MAG and Makler, M and Ogando, RLC and de Simoni, F and Mesquita, AA},
  journal={The Astronomical Journal},
  volume={142},
  number={2},
  pages={41},
  year={2011},
  publisher={The American Astronomical Society}
}

@article{zhou2026astrid,
  title={The ASTRID Simulation at z= 0: From Massive Black Holes to Large-scale Structure},
  author={Zhou, Yihao and Di Matteo, Tiziana and Bird, Simeon and Croft, Rupert and Ni, Yueying and Yang, Yanhui and Chen, Nianyi and Lachance, Patrick and Zhang, Xiaowen and Hafezianzadeh, Fatemeh},
  journal={The Astrophysical Journal},
  volume={999},
  number={1},
  pages={41},
  year={2026},
  publisher={The American Astronomical Society}
}

@ARTICLE{Donnari2019,
       author = {{Donnari}, Martina and {Pillepich}, Annalisa and {Nelson}, Dylan and {Vogelsberger}, Mark and {Genel}, Shy and {Weinberger}, Rainer and {Marinacci}, Federico and {Springel}, Volker and {Hernquist}, Lars},
        title = "{The star formation activity of IllustrisTNG galaxies: main sequence, UVJ diagram, quenched fractions, and systematics}",
      journal = {\mnras},
         year = 2019,
        month = jun,
       volume = {485},
       number = {4},
        pages = {4817-4840},
          doi = {10.1093/mnras/stz712},
archivePrefix = {arXiv},
       eprint = {1812.07584},
 primaryClass = {astro-ph.GA},
       adsurl = {https://ui.adsabs.harvard.edu/abs/2019MNRAS.485.4817D}
}

@ARTICLE{planck2015,
       author = {{Planck Collaboration} and {Ade}, P.~A.~R. and {Aghanim}, N. and {Arnaud}, M. and {Ashdown}, M. and {Aumont}, J. and {Baccigalupi}, C. and {Banday}, A.~J. and {Barreiro}, R.~B. and {Bartlett}, J.~G. and {Bartolo}, N. and {Battaner}, E. and {Battye}, R. and {Benabed}, K. and {Beno{\^\i}t}, A. and {Benoit-L{\'e}vy}, A. and {Bernard}, J.-P. and {Bersanelli}, M. and {Bielewicz}, P. and {Bock}, J.~J. and {Bonaldi}, A. and {Bonavera}, L. and {Bond}, J.~R. and {Borrill}, J. and {Bouchet}, F.~R. and {Boulanger}, F. and {Bucher}, M. and {Burigana}, C. and {Butler}, R.~C. and {Calabrese}, E. and {Cardoso}, J.-F. and {Catalano}, A. and {Challinor}, A. and {Chamballu}, A. and {Chary}, R.-R. and {Chiang}, H.~C. and {Chluba}, J. and {Christensen}, P.~R. and {Church}, S. and {Clements}, D.~L. and {Colombi}, S. and {Colombo}, L.~P.~L. and {Combet}, C. and {Coulais}, A. and {Crill}, B.~P. and {Curto}, A. and {Cuttaia}, F. and {Danese}, L. and {Davies}, R.~D. and {Davis}, R.~J. and {de Bernardis}, P. and {de Rosa}, A. and {de Zotti}, G. and {Delabrouille}, J. and {D{\'e}sert}, F.-X. and {Di Valentino}, E. and {Dickinson}, C. and {Diego}, J.~M. and {Dolag}, K. and {Dole}, H. and {Donzelli}, S. and {Dor{\'e}}, O. and {Douspis}, M. and {Ducout}, A. and {Dunkley}, J. and {Dupac}, X. and {Efstathiou}, G. and {Elsner}, F. and {En{\ss}lin}, T.~A. and {Eriksen}, H.~K. and {Farhang}, M. and {Fergusson}, J. and {Finelli}, F. and {Forni}, O. and {Frailis}, M. and {Fraisse}, A.~A. and {Franceschi}, E. and {Frejsel}, A. and {Galeotta}, S. and {Galli}, S. and {Ganga}, K. and {Gauthier}, C. and {Gerbino}, M. and {Ghosh}, T. and {Giard}, M. and {Giraud-H{\'e}raud}, Y. and {Giusarma}, E. and {Gjerl{\o}w}, E. and {Gonz{\'a}lez-Nuevo}, J. and {G{\'o}rski}, K.~M. and {Gratton}, S. and {Gregorio}, A. and {Gruppuso}, A. and {Gudmundsson}, J.~E. and {Hamann}, J. and {Hansen}, F.~K. and {Hanson}, D. and {Harrison}, D.~L. and {Helou}, G. and {Henrot-Versill{\'e}}, S. and {Hern{\'a}ndez-Monteagudo}, C. and {Herranz}, D. and {Hildebrandt}, S.~R. and {Hivon}, E. and {Hobson}, M. and {Holmes}, W.~A. and {Hornstrup}, A. and {Hovest}, W. and {Huang}, Z. and {Huffenberger}, K.~M. and {Hurier}, G. and {Jaffe}, A.~H. and {Jaffe}, T.~R. and {Jones}, W.~C. and {Juvela}, M. and {Keih{\"a}nen}, E. and {Keskitalo}, R. and {Kisner}, T.~S. and {Kneissl}, R. and {Knoche}, J. and {Knox}, L. and {Kunz}, M. and {Kurki-Suonio}, H. and {Lagache}, G. and {L{\"a}hteenm{\"a}ki}, A. and {Lamarre}, J.-M. and {Lasenby}, A. and {Lattanzi}, M. and {Lawrence}, C.~R. and {Leahy}, J.~P. and {Leonardi}, R. and {Lesgourgues}, J. and {Levrier}, F. and {Lewis}, A. and {Liguori}, M. and {Lilje}, P.~B. and {Linden-V{\o}rnle}, M. and {L{\'o}pez-Caniego}, M. and {Lubin}, P.~M. and {Mac{\'\i}as-P{\'e}rez}, J.~F. and {Maggio}, G. and {Maino}, D. and {Mandolesi}, N. and {Mangilli}, A. and {Marchini}, A. and {Maris}, M. and {Martin}, P.~G. and {Martinelli}, M. and {Mart{\'\i}nez-Gonz{\'a}lez}, E. and {Masi}, S. and {Matarrese}, S. and {McGehee}, P. and {Meinhold}, P.~R. and {Melchiorri}, A. and {Melin}, J.-B. and {Mendes}, L. and {Mennella}, A. and {Migliaccio}, M. and {Millea}, M. and {Mitra}, S. and {Miville-Desch{\^e}nes}, M.-A. and {Moneti}, A. and {Montier}, L. and {Morgante}, G. and {Mortlock}, D. and {Moss}, A. and {Munshi}, D. and {Murphy}, J.~A. and {Naselsky}, P. and {Nati}, F. and {Natoli}, P. and {Netterfield}, C.~B. and {N{\o}rgaard-Nielsen}, H.~U. and {Noviello}, F. and {Novikov}, D. and {Novikov}, I. and {Oxborrow}, C.~A. and {Paci}, F. and {Pagano}, L. and {Pajot}, F. and {Paladini}, R. and {Paoletti}, D. and {Partridge}, B. and {Pasian}, F. and {Patanchon}, G. and {Pearson}, T.~J. and {Perdereau}, O. and {Perotto}, L. and {Perrotta}, F. and {Pettorino}, V. and {Piacentini}, F. and {Piat}, M. and {Pierpaoli}, E. and {Pietrobon}, D. and {Plaszczynski}, S. and {Pointecouteau}, E. and {Polenta}, G. and {Popa}, L. and {Pratt}, G.~W. and {Pr{\'e}zeau}, G.},
        title = "{Planck 2015 results. XIII. Cosmological parameters}",
      journal = {\aap},
         year = 2016,
        month = sep,
       volume = {594},
          eid = {A13},
        pages = {A13},
          doi = {10.1051/0004-6361/201525830},
archivePrefix = {arXiv},
       eprint = {1502.01589},
 primaryClass = {astro-ph.CO},
       adsurl = {https://ui.adsabs.harvard.edu/abs/2016A&A...594A..13P}
}

@MISC{rtn-095,
       author = {{NSF-DOE Vera C. Rubin Observatory Team} and {Acero-Cuellar}, Tatiana and {Acosta}, Emily and {Adair}, Christina L. and {Adari}, Prakruth and {Adelman-McCarthy}, Jennifer K. and {Alexov}, Anastasia and {Allbery}, Russ and {Allsman}, Robyn and {AlSayyad}, Yusra and {Amado}, Jhonatan and {Amouroux}, Nathan and {Antilogus}, Pierre and {Aracena Alcayaga}, Alexis and {Aravena-Rojas}, Gonzalo and {Araya Cortes}, Claudio H. and {Aubourg}, {\'E}ric and {Axelrod}, Tim S. and {Banovetz}, John and {Barr{\'\i}a}, Carlos and {Bauer}, Amanda E. and {Bauman}, Brian J. and {Bechtol}, Ellen and {Bechtol}, Keith and {Becker}, Andrew C. and {Becker}, Valerie R. and {Beckett}, Mark G. and {Bellm}, Eric C. and {Bernardinelli}, Pedro H. and {Bianco}, Federica B. and {Blum}, Robert D. and {Bogart}, Joanne and {Bolton}, Adam and {Booth}, Michael T. and {Bosch}, James F. and {Boucaud}, Alexandre and {Boutigny}, Dominique and {Bovill}, Robert A. and {Bradshaw}, Andrew and {Bregeon}, Johan and {Brescia}, Massimo and {Brondel}, Brian J. and {Broughton}, Alex and {Budlong}, Audrey and {Buffat}, Dimitri and {Calabrese}, Daniel and {Canestrari}, Rodolfo and {Caplar}, Neven and {Carlin}, Jeffrey L. and {Ceballo}, Ross and {Chandler}, Colin Orion and {Chang}, Chihway and {Charles-Emerson}, Glenaver and {Chiang}, Hsin-Fang and {Chiang}, James and {Choi}, Yumi and {Christensen}, Eric J. and {Claver}, Charles F. and {Clements}, Andy W. and {Cockrum}, Joseph J. and {Cohen-Tanugi}, Johann and {Colleoni}, Franco and {Combet}, C{\'e}line and {Connolly}, Andrew J. and {Constanzo C{\'o}rdova}, Julio Eduardo and {Contreras}, Hans E. and {Crenshaw}, John Franklin and {Dagoret-Campagne}, Sylvie and {Daniel}, Scott F. and {Daruich}, Felipe and {Daubard}, Guillaume and {Daues}, Greg and {Dennihy}, Erik and {Deppe}, Stephanie J.~H. and {Digel}, Seth W. and {Doherty}, Peter E. and {Drlica-Wagner}, Alex and {Dubois-Felsmann}, Gregory P. and {Economou}, Frossie and {Eiger}, Orion and {Eisert}, Lukas and {Eisner}, Alan M. and {Englert}, Anthony and {Erb}, Baden and {Fabrega}, Juan A. and {Fagrelius}, Parker and {Fanning}, Kevin and {Fausti Neto}, Angelo and {Ferguson}, Peter S. and {Fert{\'e}}, Agn{\`e}s and {Fisher-Levine}, Merlin and {Fonseca Alvarez}, Gloria and {Foss}, Michael D. and {Fouchez}, Dominique and {Fuchs}, Dan C. and {Gangler}, Emmanuel and {Gaponenko}, Igor and {Garcia}, Julen and {Gates}, John H. and {Gill}, Ranpal K. and {Giro}, Enrico and {Glanzman}, Thomas and {Godoy}, Robinson and {Goodenow}, Iain and {Gorsuch}, Miranda R. and {Gower}, Michelle and {Granvik}, Mikael and {Greenstreet}, Sarah and {Guan}, Wen and {Guillemin}, Thibault and {Guy}, Leanne P. and {Hascall}, Diane and {Hascall}, Patrick A. and {Heinze}, Aren Nathaniel and {Hernandez}, Fabio and {Herner}, Kenneth and {Herrold}, Ardis and {Higgs}, Clare R. and {Hoblitt}, Joshua and {Howard}, Erin Leigh and {Hyun}, Minhee and {Ingraham}, Patrick and {Irving}, David H. and {Ivezi{\'c}}, {\v{Z}}eljko and {Jacoby}, Suzanne H. and {Jannuzi}, Buell T. and {Jarugula}, Sreevani and {Jee}, M. James and {Jenness}, Tim and {Jennings}, Toby C. and {Jeremie}, Andrea and {Jernigan}, Garrett and {Jim{\'e}nez Mej{\'\i}as}, David and {Johnson}, Anthony S. and {Jones}, R. Lynne and {Jones}, Roger William Lewis and {Juramy-Gilles}, Claire and {Juri{\'c}}, Mario and {Kahn}, Steven M. and {Kalmbach}, J. Bryce and {Kang}, Yijung and {Kannawadi}, Arun and {Kantor}, Jeffrey P. and {Karavakis}, Edward and {Kelkar}, Kshitija and {Kelvin}, Lee S. and {Kotov}, Ivan V. and {Kov{\'a}cs}, G{\'a}bor and {Kowalik}, Mikolaj and {Krabbendam}, Victor L. and {Krughoff}, K. Simon and {Kub{\'a}nek}, Petr and {Kurlander}, Jacob A. and {Kusulja}, Mile and {Lage}, Craig S. and {Lago}, Paulo J.~A. and {Laliotis}, Katherine and {Lange}, Travis and {Laporte}, Didier and {Lau}, Ryan M. and {Lazarte}, Juan Carlos and {Le Boulc'h}, Quentin and {L{\'e}get}, Pierre-Fran{\c{c}}ois and {Le Guillou}, Laurent and {Levine}, Benjamin and {Liang}, Ming and {Liang}, Shuang and {Lim}, Kian-Tat and {von der Linden}, Anja and {Lin}, Huan and {Lopez}, Margaux and {Lopez Toro}, Juan J. and {Love}, Peter and {Lupton}, Robert H. and {Lust}, Nate B. and {MacArthur}, Lauren A. and {MacBride}, Sean Patrick and {Madejski}, Greg M. and {Mainetti}, Gabriele and {Margheim}, Steven J. and {Markiewicz}, Thomas W. and {Marshall}, Phil and {Marshall}, Stuart and {Maulen}, Guido and {May}, Morgan and {McCormick}, Jeremy and {McKay}, David and {McKercher}, Robert and {Megias Homar}, Guillem and {Meisner}, Aaron M. and {Menanteau}, Felipe and {Mentzer}, Heather R. and {Metzger}, Kristen and {Meyers}, Joshua E. and {Miller}, Michelle and {Mills}, David J. and {Moeyens}, Joachim and {Moniez}, Marc and {Moolekamp}, Fred E. and {Morales Mar{\'\i}n}, C.~A.~L.},
        title = "{RTN-095: The Vera C. Rubin Observatory Data Preview 1}",
 howpublished = {NSF-DOE Vera C. Rubin Observatory Technical Report},
         year = 2025,
        month = jan,
        pages = {31},
          doi = {10.71929/RUBIN/2570536},
       adsurl = {https://ui.adsabs.harvard.edu/abs/2025rubn.rept...31N}
}

@ARTICLE{crenshaw2024,
       author = {{Crenshaw}, John Franklin and {Kalmbach}, J. Bryce and {Gagliano}, Alexander and {Yan}, Ziang and {Connolly}, Andrew J. and {Malz}, Alex I. and {Schmidt}, Samuel J. and {The LSST Dark Energy Science Collaboration}},
        title = "{Probabilistic Forward Modeling of Galaxy Catalogs with Normalizing Flows}",
      journal = {\aj},
         year = 2024,
        month = aug,
       volume = {168},
       number = {2},
          eid = {80},
        pages = {80},
          doi = {10.3847/1538-3881/ad54bf},
archivePrefix = {arXiv},
       eprint = {2405.04740},
 primaryClass = {astro-ph.IM},
       adsurl = {https://ui.adsabs.harvard.edu/abs/2024AJ....168...80C}
}

@ARTICLE{Ivezic2019,
       author = {{Ivezi{\'c}}, {\v{Z}}eljko and {Kahn}, Steven M. and {Tyson}, J. Anthony and {Abel}, Bob and {Acosta}, Emily and {Allsman}, Robyn and {Alonso}, David and {AlSayyad}, Yusra and {Anderson}, Scott F. and {Andrew}, John and {Angel}, James Roger P. and {Angeli}, George Z. and {Ansari}, Reza and {Antilogus}, Pierre and {Araujo}, Constanza and {Armstrong}, Robert and {Arndt}, Kirk T. and {Astier}, Pierre and {Aubourg}, {\'E}ric and {Auza}, Nicole and {Axelrod}, Tim S. and {Bard}, Deborah J. and {Barr}, Jeff D. and {Barrau}, Aurelian and {Bartlett}, James G. and {Bauer}, Amanda E. and {Bauman}, Brian J. and {Baumont}, Sylvain and {Bechtol}, Ellen and {Bechtol}, Keith and {Becker}, Andrew C. and {Becla}, Jacek and {Beldica}, Cristina and {Bellavia}, Steve and {Bianco}, Federica B. and {Biswas}, Rahul and {Blanc}, Guillaume and {Blazek}, Jonathan and {Blandford}, Roger D. and {Bloom}, Josh S. and {Bogart}, Joanne and {Bond}, Tim W. and {Booth}, Michael T. and {Borgland}, Anders W. and {Borne}, Kirk and {Bosch}, James F. and {Boutigny}, Dominique and {Brackett}, Craig A. and {Bradshaw}, Andrew and {Brandt}, William Nielsen and {Brown}, Michael E. and {Bullock}, James S. and {Burchat}, Patricia and {Burke}, David L. and {Cagnoli}, Gianpietro and {Calabrese}, Daniel and {Callahan}, Shawn and {Callen}, Alice L. and {Carlin}, Jeffrey L. and {Carlson}, Erin L. and {Chandrasekharan}, Srinivasan and {Charles-Emerson}, Glenaver and {Chesley}, Steve and {Cheu}, Elliott C. and {Chiang}, Hsin-Fang and {Chiang}, James and {Chirino}, Carol and {Chow}, Derek and {Ciardi}, David R. and {Claver}, Charles F. and {Cohen-Tanugi}, Johann and {Cockrum}, Joseph J. and {Coles}, Rebecca and {Connolly}, Andrew J. and {Cook}, Kem H. and {Cooray}, Asantha and {Covey}, Kevin R. and {Cribbs}, Chris and {Cui}, Wei and {Cutri}, Roc and {Daly}, Philip N. and {Daniel}, Scott F. and {Daruich}, Felipe and {Daubard}, Guillaume and {Daues}, Greg and {Dawson}, William and {Delgado}, Francisco and {Dellapenna}, Alfred and {de Peyster}, Robert and {de Val-Borro}, Miguel and {Digel}, Seth W. and {Doherty}, Peter and {Dubois}, Richard and {Dubois-Felsmann}, Gregory P. and {Durech}, Josef and {Economou}, Frossie and {Eifler}, Tim and {Eracleous}, Michael and {Emmons}, Benjamin L. and {Fausti Neto}, Angelo and {Ferguson}, Henry and {Figueroa}, Enrique and {Fisher-Levine}, Merlin and {Focke}, Warren and {Foss}, Michael D. and {Frank}, James and {Freemon}, Michael D. and {Gangler}, Emmanuel and {Gawiser}, Eric and {Geary}, John C. and {Gee}, Perry and {Geha}, Marla and {Gessner}, Charles J.~B. and {Gibson}, Robert R. and {Gilmore}, D. Kirk and {Glanzman}, Thomas and {Glick}, William and {Goldina}, Tatiana and {Goldstein}, Daniel A. and {Goodenow}, Iain and {Graham}, Melissa L. and {Gressler}, William J. and {Gris}, Philippe and {Guy}, Leanne P. and {Guyonnet}, Augustin and {Haller}, Gunther and {Harris}, Ron and {Hascall}, Patrick A. and {Haupt}, Justine and {Hernandez}, Fabio and {Herrmann}, Sven and {Hileman}, Edward and {Hoblitt}, Joshua and {Hodgson}, John A. and {Hogan}, Craig and {Howard}, James D. and {Huang}, Dajun and {Huffer}, Michael E. and {Ingraham}, Patrick and {Innes}, Walter R. and {Jacoby}, Suzanne H. and {Jain}, Bhuvnesh and {Jammes}, Fabrice and {Jee}, M. James and {Jenness}, Tim and {Jernigan}, Garrett and {Jevremovi{\'c}}, Darko and {Johns}, Kenneth and {Johnson}, Anthony S. and {Johnson}, Margaret W.~G. and {Jones}, R. Lynne and {Juramy-Gilles}, Claire and {Juri{\'c}}, Mario and {Kalirai}, Jason S. and {Kallivayalil}, Nitya J. and {Kalmbach}, Bryce and {Kantor}, Jeffrey P. and {Karst}, Pierre and {Kasliwal}, Mansi M. and {Kelly}, Heather and {Kessler}, Richard and {Kinnison}, Veronica and {Kirkby}, David and {Knox}, Lloyd and {Kotov}, Ivan V. and {Krabbendam}, Victor L. and {Krughoff}, K. Simon and {Kub{\'a}nek}, Petr and {Kuczewski}, John and {Kulkarni}, Shri and {Ku}, John and {Kurita}, Nadine R. and {Lage}, Craig S. and {Lambert}, Ron and {Lange}, Travis and {Langton}, J. Brian and {Le Guillou}, Laurent and {Levine}, Deborah and {Liang}, Ming and {Lim}, Kian-Tat and {Lintott}, Chris J. and {Long}, Kevin E. and {Lopez}, Margaux and {Lotz}, Paul J. and {Lupton}, Robert H. and {Lust}, Nate B. and {MacArthur}, Lauren A. and {Mahabal}, Ashish and {Mandelbaum}, Rachel and {Markiewicz}, Thomas W. and {Marsh}, Darren S. and {Marshall}, Philip J. and {Marshall}, Stuart and {May}, Morgan and {McKercher}, Robert and {McQueen}, Michelle and {Meyers}, Joshua and {Migliore}, Myriam and {Miller}, Michelle and {Mills}, David J.},
        title = "{LSST: From Science Drivers to Reference Design and Anticipated Data Products}",
      journal = {\apj},
         year = 2019,
        month = mar,
       volume = {873},
       number = {2},
          eid = {111},
        pages = {111},
          doi = {10.3847/1538-4357/ab042c},
archivePrefix = {arXiv},
       eprint = {0805.2366},
 primaryClass = {astro-ph},
       adsurl = {https://ui.adsabs.harvard.edu/abs/2019ApJ...873..111I}
}

@article{10.1093/mnras/staa3802,
    author = {Fortuna, Maria Cristina and Hoekstra, Henk and Joachimi, Benjamin and Johnston, Harry and Chisari, Nora Elisa and Georgiou, Christos and Mahony, Constance},
    title = {The halo model as a versatile tool to predict intrinsic alignments},
    journal = {Monthly Notices of the Royal Astronomical Society},
    volume = {501},
    number = {2},
    pages = {2983-3002},
    year = {2021},
    month = {02},
    issn = {0035-8711},
    doi = {10.1093/mnras/staa3802},
    url = {https://doi.org/10.1093/mnras/staa3802},
    eprint = {https://academic.oup.com/mnras/article-pdf/501/2/2983/35559288/staa3802.pdf},
}

@article{10.1093/mnras/stv2154,
    author = {Chisari, N. and Codis, S. and Laigle, C. and Dubois, Y. and Pichon, C. and Devriendt, J. and Slyz, A. and Miller, L. and Gavazzi, R. and Benabed, K.},
    title = {Intrinsic alignments of galaxies in the Horizon-AGN cosmological hydrodynamical simulation},
    journal = {Monthly Notices of the Royal Astronomical Society},
    volume = {454},
    number = {3},
    pages = {2736-2753},
    year = {2015},
    month = {12},
    issn = {0035-8711},
    doi = {10.1093/mnras/stv2154},
    url = {https://doi.org/10.1093/mnras/stv2154},
    eprint = {https://academic.oup.com/mnras/article-pdf/454/3/2736/4036372/stv2154.pdf},
}

@ARTICLE{2024arXiv241022272M,
       author = {{McCullough}, J. and {Amon}, A. and {Legnani}, E. and {Gruen}, D. and {Roodman}, A. and {Friedrich}, O. and {MacCrann}, N. and {Becker}, M.~R. and {Myles}, J. and {Dodelson}, S. and {Samuroff}, S. and {Blazek}, J. and {Prat}, J. and {Honscheid}, K. and {Pieres}, A. and {Fert{\'e}}, A. and {Alarcon}, A. and {Drlica-Wagner}, A. and {Choi}, A. and {Navarro-Alsina}, A. and {Campos}, A. and {Plazas Malag{\'o}n}, A.~A. and {Porredon}, A. and {Farahi}, A. and {Ross}, A.~J. and {Carnero Rosell}, A. and {Yin}, B. and {Flaugher}, B. and {Yanny}, B. and {S{\'a}nchez}, C. and {Chang}, C. and {Davis}, C. and {To}, C. and {Doux}, C. and {Brooks}, D. and {James}, D.~J. and {Sanchez Cid}, D. and {Hollowood}, D.~L. and {Huterer}, D. and {Rykoff}, E.~S. and {Gaztanaga}, E. and {Huff}, E.~M. and {Suchyta}, E. and {Sheldon}, E. and {Sanchez}, E. and {Tarsitano}, F. and {Andrade-Oliveira}, F. and {Castander}, F.~J. and {Bernstein}, G.~M. and {Gutierrez}, G. and {Giannini}, G. and {Tarle}, G. and {Diehl}, H.~T. and {Huang}, H. and {Harrison}, I. and {Sevilla-Noarbe}, I. and {Tutusaus}, I. and {Ferrero}, I. and {Elvin-Poole}, J. and {Marshall}, J.~L. and {Muir}, J. and {Weller}, J. and {Zuntz}, J. and {Carretero}, J. and {DeRose}, J. and {Frieman}, J. and {Cordero}, J. and {De Vicente}, J. and {Garc{\'\i}a-Bellido}, J. and {Mena-Fern{\'a}ndez}, J. and {Eckert}, K. and {Romer}, A.~K. and {Bechtol}, K. and {Herner}, K. and {Kuehn}, K. and {Secco}, L.~F. and {da Costa}, L.~N. and {Paterno}, M. and {Soares-Santos}, 21 M. and {Gatti}, M. and {Raveri}, M. and {Yamamoto}, M. and {Smith}, M. and {Carrasco Kind}, M. and {Troxel}, M.~A. and {Aguena}, M. and {Jarvis}, M. and {Swanson}, M.~E.~C. and {Weaverdyck}, N. and {Lahav}, O. and {Doel}, P. and {Wiseman}, P. and {Miquel}, R. and {Gruendl}, R.~A. and {Cawthon}, R. and {Allam}, S. and {Hinton}, S.~R. and {Bridle}, S.~L. and {Bocquet}, S. and {Desai}, S. and {Pandey}, S. and {Everett}, S. and {Lee}, S. and {Shin}, T. and {Palmese}, A. and {Conselice}, C. and {Burke}, D.~L. and {Buckley-Geer}, E. and {Lima}, M. and {Vincenzi}, M. and {Pereira}, M.~E.~S. and {Crocce}, M. and {Schubnell}, M. and {Jeffrey}, N. and {Alves}, O. and {Vikram}, V. and {Zhang}, Y. and {DES Collaboration}},
        title = "{Dark Energy Survey Year 3: Blue Shear}",
      journal = {arXiv e-prints},
         year = 2024,
        month = oct,
          eid = {arXiv:2410.22272},
        pages = {arXiv:2410.22272},
          doi = {10.48550/arXiv.2410.22272},
archivePrefix = {arXiv},
       eprint = {2410.22272},
 primaryClass = {astro-ph.CO},
       adsurl = {https://ui.adsabs.harvard.edu/abs/2024arXiv241022272M}
}

@ARTICLE{2013MNRAS.431..477J,
       author = {{Joachimi}, B. and {Semboloni}, E. and {Bett}, P.~E. and {Hartlap}, J. and {Hilbert}, S. and {Hoekstra}, H. and {Schneider}, P. and {Schrabback}, T.},
        title = "{Intrinsic galaxy shapes and alignments - I. Measuring and modelling COSMOS intrinsic galaxy ellipticities}",
      journal = {\mnras},
         year = 2013,
        month = may,
       volume = {431},
       number = {1},
        pages = {477-492},
          doi = {10.1093/mnras/stt172},
archivePrefix = {arXiv},
       eprint = {1203.6833},
 primaryClass = {astro-ph.CO},
       adsurl = {https://ui.adsabs.harvard.edu/abs/2013MNRAS.431..477J}
}

@ARTICLE{2025AA...699A.252G,
       author = {{Georgiou}, Christos and {Chisari}, Nora Elisa and {Bilicki}, Maciej and {La Barbera}, Francesco and {Napolitano}, Nicola R. and {Roy}, Nivya and {Tortora}, Crescenzo},
        title = "{Intrinsic galaxy alignments in the KiDS-1000 bright sample: Dependence on colour, luminosity, morphology, and galaxy scale}",
      journal = {\aap},
         year = 2025,
        month = jul,
       volume = {699},
          eid = {A252},
        pages = {A252},
          doi = {10.1051/0004-6361/202554134},
archivePrefix = {arXiv},
       eprint = {2502.09452},
 primaryClass = {astro-ph.CO},
       adsurl = {https://ui.adsabs.harvard.edu/abs/2025A&A...699A.252G}
}

@ARTICLE{2023MNRAS.523.5899D,
       author = {{Delgado}, Ana Maria and {Hadzhiyska}, Boryana and {Bose}, Sownak and {Springel}, Volker and {Hernquist}, Lars and {Barrera}, Monica and {Pakmor}, R{\"u}diger and {Ferlito}, Fulvio and {Kannan}, Rahul and {Hern{\'a}ndez-Aguayo}, C{\'e}sar and {White}, Simon D.~M. and {Frenk}, Carlos},
        title = "{The MillenniumTNG project: intrinsic alignments of galaxies and haloes}",
      journal = {\mnras},
         year = 2023,
        month = aug,
       volume = {523},
       number = {4},
        pages = {5899-5914},
          doi = {10.1093/mnras/stad1781},
archivePrefix = {arXiv},
       eprint = {2304.12346},
 primaryClass = {astro-ph.CO},
       adsurl = {https://ui.adsabs.harvard.edu/abs/2023MNRAS.523.5899D}
}

@article{10.1093/mnras/stac1424,
    author = {Jagvaral, Yesukhei and Singh, Sukhdeep and Mandelbaum, Rachel},
    title = {Intrinsic alignments of bulges and discs},
    journal = {Monthly Notices of the Royal Astronomical Society},
    volume = {514},
    number = {1},
    pages = {1021-1033},
    year = {2022},
    month = {07},
    issn = {0035-8711},
    doi = {10.1093/mnras/stac1424},
    url = {https://doi.org/10.1093/mnras/stac1424},
    eprint = {https://academic.oup.com/mnras/article-pdf/514/1/1021/43984851/stac1424.pdf},
}

@article{Nelson_2017,
   title={First results from the IllustrisTNG simulations: the galaxy colour bimodality},
   volume={475},
   ISSN={1365-2966},
   url={http://dx.doi.org/10.1093/mnras/stx3040},
   DOI={10.1093/mnras/stx3040},
   number={1},
   journal={Monthly Notices of the Royal Astronomical Society},
   publisher={Oxford University Press (OUP)},
   author={Nelson, Dylan and Pillepich, Annalisa and Springel, Volker and Weinberger, Rainer and Hernquist, Lars and Pakmor, Rüdiger and Genel, Shy and Torrey, Paul and Vogelsberger, Mark and Kauffmann, Guinevere and Marinacci, Federico and Naiman, Jill},
   year={2017},
   month=Nov, pages={624–647} }

@article{Tenneti_2016,
   title={Intrinsic alignments of disc and elliptical galaxies in the MassiveBlack-II and Illustris simulations},
   volume={462},
   ISSN={1365-2966},
   url={http://dx.doi.org/10.1093/mnras/stw1823},
   DOI={10.1093/mnras/stw1823},
   number={3},
   journal={Monthly Notices of the Royal Astronomical Society},
   publisher={Oxford University Press (OUP)},
   author={Tenneti, Ananth and Mandelbaum, Rachel and Di Matteo, Tiziana},
   year={2016},
   month=July, pages={2668–2680} }

@article{Samuroff_2021,
   title={Advances in constraining intrinsic alignment models with hydrodynamic simulations},
   volume={508},
   ISSN={1365-2966},
   url={http://dx.doi.org/10.1093/mnras/stab2520},
   DOI={10.1093/mnras/stab2520},
   number={1},
   journal={Monthly Notices of the Royal Astronomical Society},
   publisher={Oxford University Press (OUP)},
   author={Samuroff, S and Mandelbaum, R and Blazek, J},
   year={2021},
   month=Sept, pages={637–664} }

@article{Conselice_2014,
   title={The Evolution of Galaxy Structure Over Cosmic Time},
   volume={52},
   ISSN={1545-4282},
   url={http://dx.doi.org/10.1146/annurev-astro-081913-040037},
   DOI={10.1146/annurev-astro-081913-040037},
   number={1},
   journal={Annual Review of Astronomy and Astrophysics},
   publisher={Annual Reviews},
   author={Conselice, Christopher J.},
   year={2014},
   month=Aug, pages={291–337} }

@article{Secco_2022,
   title={Dark Energy Survey Year 3 results: Cosmology from cosmic shear and robustness to modeling uncertainty},
   volume={105},
   ISSN={2470-0029},
   url={http://dx.doi.org/10.1103/PhysRevD.105.023515},
   DOI={10.1103/physrevd.105.023515},
   number={2},
   journal={Physical Review D},
   publisher={American Physical Society (APS)},
   author={Secco, L. F. and Samuroff, S. and Krause, E. and Jain, B. and Blazek, J. and Raveri, M. and Campos, A. and Amon, A. and Chen, A. and Doux, C. and Choi, A. and Gruen, D. and Bernstein, G. M. and Chang, C. and DeRose, J. and Myles, J. and Ferté, A. and Lemos, P. and Huterer, D. and Prat, J. and Troxel, M. A. and MacCrann, N. and Liddle, A. R. and Kacprzak, T. and Fang, X. and Sánchez, C. and Pandey, S. and Dodelson, S. and Chintalapati, P. and Hoffmann, K. and Alarcon, A. and Alves, O. and Andrade-Oliveira, F. and Baxter, E. J. and Bechtol, K. and Becker, M. R. and Brandao-Souza, A. and Camacho, H. and Carnero Rosell, A. and Carrasco Kind, M. and Cawthon, R. and Cordero, J. P. and Crocce, M. and Davis, C. and Di Valentino, E. and Drlica-Wagner, A. and Eckert, K. and Eifler, T. F. and Elidaiana, M. and Elsner, F. and Elvin-Poole, J. and Everett, S. and Fosalba, P. and Friedrich, O. and Gatti, M. and Giannini, G. and Gruendl, R. A. and Harrison, I. and Hartley, W. G. and Herner, K. and Huang, H. and Huff, E. M. and Jarvis, M. and Jeffrey, N. and Kuropatkin, N. and Leget, P.-F. and Muir, J. and Mccullough, J. and Navarro Alsina, A. and Omori, Y. and Park, Y. and Porredon, A. and Rollins, R. and Roodman, A. and Rosenfeld, R. and Ross, A. J. and Rykoff, E. S. and Sanchez, J. and Sevilla-Noarbe, I. and Sheldon, E. S. and Shin, T. and Troja, A. and Tutusaus, I. and Varga, T. N. and Weaverdyck, N. and Wechsler, R. H. and Yanny, B. and Yin, B. and Zhang, Y. and Zuntz, J. and Abbott, T. M. C. and Aguena, M. and Allam, S. and Annis, J. and Bacon, D. and Bertin, E. and Bhargava, S. and Bridle, S. L. and Brooks, D. and Buckley-Geer, E. and Burke, D. L. and Carretero, J. and Costanzi, M. and da Costa, L. N. and De Vicente, J. and Diehl, H. T. and Dietrich, J. P. and Doel, P. and Ferrero, I. and Flaugher, B. and Frieman, J. and García-Bellido, J. and Gaztanaga, E. and Gerdes, D. W. and Giannantonio, T. and Gschwend, J. and Gutierrez, G. and Hinton, S. R. and Hollowood, D. L. and Honscheid, K. and Hoyle, B. and James, D. J. and Jeltema, T. and Kuehn, K. and Lahav, O. and Lima, M. and Lin, H. and Maia, M. A. G. and Marshall, J. L. and Martini, P. and Melchior, P. and Menanteau, F. and Miquel, R. and Mohr, J. J. and Morgan, R. and Ogando, R. L. C. and Palmese, A. and Paz-Chinchón, F. and Petravick, D. and Pieres, A. and Plazas Malagón, A. A. and Rodriguez-Monroy, M. and Romer, A. K. and Sanchez, E. and Scarpine, V. and Schubnell, M. and Scolnic, D. and Serrano, S. and Smith, M. and Soares-Santos, M. and Suchyta, E. and Swanson, M. E. C. and Tarle, G. and Thomas, D. and To, C. and },
   year={2022},
   month=Jan }

@article{Krause_2015,
   title={The impact of intrinsic alignment on current and future cosmic shear surveys},
   volume={456},
   ISSN={1365-2966},
   url={http://dx.doi.org/10.1093/mnras/stv2615},
   DOI={10.1093/mnras/stv2615},
   number={1},
   journal={Monthly Notices of the Royal Astronomical Society},
   publisher={Oxford University Press (OUP)},
   author={Krause, Elisabeth and Eifler, Tim and Blazek, Jonathan},
   year={2015},
   month=Dec, pages={207–222} }

@article{Somerville_2015,
   title={Physical Models of Galaxy Formation in a Cosmological Framework},
   volume={53},
   ISSN={1545-4282},
   url={http://dx.doi.org/10.1146/annurev-astro-082812-140951},
   DOI={10.1146/annurev-astro-082812-140951},
   number={1},
   journal={Annual Review of Astronomy and Astrophysics},
   publisher={Annual Reviews},
   author={Somerville, Rachel S. and Davé, Romeel},
   year={2015},
   month=Aug, pages={51–113} }

@article{10.1093/mnras/stz2197,
    author = {Samuroff, S and Blazek, J and Troxel, M A and MacCrann, N and Krause, E and Leonard, C D and Prat, J and Gruen, D and Dodelson, S and Eifler, T F and Gatti, M and Hartley, W G and Hoyle, B and Larsen, P and Zuntz, J and Abbott, T M C and Allam, S and Annis, J and Bernstein, G M and Bertin, E and Bridle, S L and Brooks, D and Carnero Rosell, A and Carrasco Kind, M and Carretero, J and Castander, F J and Cunha, C E and da Costa, L N and Davis, C and De Vicente, J and DePoy, D L and Desai, S and Diehl, H T and Dietrich, J P and Doel, P and Flaugher, B and Fosalba, P and Frieman, J and García-Bellido, J and Gaztanaga, E and Gerdes, D W and Gruendl, R A and Gschwend, J and Gutierrez, G and Hollowood, D L and Honscheid, K and James, D J and Kuehn, K and Kuropatkin, N and Lima, M and Maia, M A G and March, M and Marshall, J L and Martini, P and Melchior, P and Menanteau, F and Miller, C J and Miquel, R and Ogando, R L C and Plazas, A A and Sanchez, E and Scarpine, V and Schindler, R and Schubnell, M and Serrano, S and Sevilla-Noarbe, I and Sheldon, E and Smith, M and Sobreira, F and Suchyta, E and Tarle, G and Thomas, D and Vikram, V and (DES Collaboration)},
    title = {Dark Energy Survey Year 1 results: constraints on intrinsic alignments and their colour dependence from galaxy clustering and weak lensing},
    journal = {Monthly Notices of the Royal Astronomical Society},
    volume = {489},
    number = {4},
    pages = {5453-5482},
    year = {2019},
    month = {11},
    issn = {0035-8711},
    doi = {10.1093/mnras/stz2197},
    url = {https://doi.org/10.1093/mnras/stz2197},
    eprint = {https://academic.oup.com/mnras/article-pdf/489/4/5453/30080335/stz2197.pdf},
}

@article{openuniverse2025,
  title={OpenUniverse2024: a shared, simulated view of the sky for the next generation of cosmological surveys},
  author={Alarcon, A and Aldoroty, L and Beltz-Mohrmann, G and Bera, A and Blazek, J and Bogart, J and Braeunlich, G and Broughton, A and Cao, K and others},
  journal={Monthly Notices of the Royal Astronomical Society},
  volume={544},
  number={4},
  pages={3799--3823},
  year={2025},
  publisher={Oxford University Press}
}

@article{to2024buzzard,
  title={Buzzard to Cardinal: Improved Mock Catalogs for Large Galaxy Surveys},
  author={To, Chun-Hao and DeRose, Joseph and Wechsler, Risa H and Rykoff, Eli and Wu, Hao-Yi and Adhikari, Susmita and Krause, Elisabeth and Rozo, Eduardo and Weinberg, David H},
  journal={The Astrophysical Journal},
  volume={961},
  number={1},
  pages={59},
  year={2024},
  publisher={The American Astronomical Society}
}

@article{team2026vera,
  title={The Vera C. Rubin Observatory Data Preview 1},
  author={Acero Cuellar, Tatiana and Acosta, Emily and Adair, Christina L and Adari, Prakruth and Adelman McCarthy, Jennifer K and Alexov, Anastasia and Allbery, Russ and Allsman, Robyn and AlSayyad, Yusra and Amado, Jhonatan and others},
  journal={arXiv e-prints},
  pages={arXiv--2603},
  year={2026}
}

@article{DeRose2019,
  title={The buzzard flock: Dark energy survey synthetic sky catalogs},
  author={DeRose, Joseph and Wechsler, Risa H and Becker, Matthew R and Busha, Michael T and Rykoff, Eli S and MacCrann, Niall and Erickson, Brandon and Evrard, August E and Kravtsov, Andrey and Gruen, Daniel and others},
  journal={arXiv preprint arXiv:1901.02401},
  year={2019}
}

@article{korytov2019,
  title={CosmoDC2: A synthetic sky catalog for dark energy science with LSST},
  author={Korytov, Danila and Hearin, Andrew and Kovacs, Eve and Larsen, Patricia and Rangel, Esteban and Hollowed, Joseph and Benson, Andrew J and Heitmann, Katrin and Mao, Yao-Yuan and Bahmanyar, Anita and others},
  journal={The Astrophysical Journal Supplement Series},
  volume={245},
  number={2},
  pages={26},
  year={2019},
  publisher={The American Astronomical Society}
}

@article{blanchard2020euclid,
  title={Euclid preparation-VII. Forecast validation for Euclid cosmological probes},
  author={Blanchard, Alain and Camera, S and Carbone, Carmelita and Cardone, VF and Casas, S and Clesse, S{\'e}bastien and Ili{\'c}, S and Kilbinger, M and Kitching, T and Kunz, Martin and others},
  journal={Astronomy \& Astrophysics},
  volume={642},
  pages={A191},
  year={2020},
  publisher={EDP sciences}
}

@article{potter2017pkdgrav3,
  title={PKDGRAV3: beyond trillion particle cosmological simulations for the next era of galaxy surveys},
  author={Potter, Douglas and Stadel, Joachim and Teyssier, Romain},
  journal={Computational Astrophysics and Cosmology},
  volume={4},
  number={1},
  pages={2},
  year={2017},
  publisher={Springer}
}

@software{peter_yoachim_2025_15832326,
  author       = {Peter Yoachim},
  title        = {lsst-sims/sims\_featureScheduler\_runs5.0: Initial
                   Release
                  },
  month        = jul,
  year         = 2025,
  publisher    = {Zenodo},
  version      = {1.0},
  doi          = {10.5281/zenodo.15832326},
  url          = {https://doi.org/10.5281/zenodo.15832326},
  swhid        = {swh:1:dir:0c8d0f83749525f88c25d27992bd53cfdc1616d3
                   ;origin=https://doi.org/10.5281/zenodo.15832325;vi
                   sit=swh:1:snp:faddc44568503fd5e82e7a8c33944dd65ad8
                   5059;anchor=swh:1:rel:60309c63e64174b8254c307aca23
                   b4f016867bb0;path=lsst-sims-
                   sims\_featureScheduler\_runs5.0-0294114
                  },
}

@ARTICLE{2022ApJS..258...11W,
       author = {{Weaver}, J.~R. and {Kauffmann}, O.~B. and {Ilbert}, O. and {McCracken}, H.~J. and {Moneti}, A. and {Toft}, S. and {Brammer}, G. and {Shuntov}, M. and {Davidzon}, I. and {Hsieh}, B.~C. and {Laigle}, C. and {Anastasiou}, A. and {Jespersen}, C.~K. and {Vinther}, J. and {Capak}, P. and {Casey}, C.~M. and {McPartland}, C.~J.~R. and {Milvang-Jensen}, B. and {Mobasher}, B. and {Sanders}, D.~B. and {Zalesky}, L. and {Arnouts}, S. and {Aussel}, H. and {Dunlop}, J.~S. and {Faisst}, A. and {Franx}, M. and {Furtak}, L.~J. and {Fynbo}, J.~P.~U. and {Gould}, K.~M.~L. and {Greve}, T.~R. and {Gwyn}, S. and {Kartaltepe}, J.~S. and {Kashino}, D. and {Koekemoer}, A.~M. and {Kokorev}, V. and {Le F{\`e}vre}, O. and {Lilly}, S. and {Masters}, D. and {Magdis}, G. and {Mehta}, V. and {Peng}, Y. and {Riechers}, D.~A. and {Salvato}, M. and {Sawicki}, M. and {Scarlata}, C. and {Scoville}, N. and {Shirley}, R. and {Silverman}, J.~D. and {Sneppen}, A. and {Smolc̆i{\'c}}, V. and {Steinhardt}, C. and {Stern}, D. and {Tanaka}, M. and {Taniguchi}, Y. and {Teplitz}, H.~I. and {Vaccari}, M. and {Wang}, W.-H. and {Zamorani}, G.},
        title = "{COSMOS2020: A Panchromatic View of the Universe to z{\ensuremath{\sim}}10 from Two Complementary Catalogs}",
      journal = {\apjs},
         year = 2022,
        month = jan,
       volume = {258},
       number = {1},
          eid = {11},
        pages = {11},
          doi = {10.3847/1538-4365/ac3078},
archivePrefix = {arXiv},
       eprint = {2110.13923},
 primaryClass = {astro-ph.GA},
       adsurl = {https://ui.adsabs.harvard.edu/abs/2022ApJS..258...11W}
}
\bibliographystyle{aasjournal}

\end{document}